\documentclass[fleqn,usenatbib]{mnras}

\usepackage{newtxtext,newtxmath}
 
\usepackage[T1]{fontenc}

\DeclareRobustCommand{\VAN}[3]{#2}
\let\VANthebibliography\thebibliography
\def\thebibliography{\DeclareRobustCommand{\VAN}[3]{##3}\VANthebibliography}

\usepackage{graphicx}	
\usepackage{amsmath}	
\usepackage{amssymb}	
\usepackage{nicefrac}

\newcommand{\gcm}{g cm$^{-2}$}
\newcommand{\medd}{\ensuremath{\dot{m}_{\mathrm{Edd}}}}
\newcommand{\MeVnuc}{\ensuremath{\mathrm{MeV nuc}^{-1}}}
\newcommand{\XTE}{XTE J1814--338}
\newcommand{\IGR}{IGR J17480--2446}
\newcommand{\msun}{\ensuremath{\mathrm{M}_{\odot}}}
\newcommand{\h}{\nicefrac12}

\title[Hotspot effect on X-ray burst ignition location]{X-ray burst ignition location on the surface of accreting X-ray pulsars: Can bursts preferentially ignite at the hotspot? }

\author[A. J. Goodwin et al.]{
A. J. Goodwin,$^{1,2,3}$\thanks{E-mail: ajgoodwin.astro@gmail.com}
A. Heger,$^{1,2,4,5}$
F. R. N. Chambers,$^{6}$
A. L. Watts,$^{6}$
and
Y. Cavecchi$^{2,7}$
\\
$^{1}$School of Physics and Astronomy, Monash University, Clayton, 3800, Australia\\
$^{2}$Joint Institute for Nuclear Astrophysics, 1 Cyclotron Laboratory, National Superconducting Cyclotron Laboratory,\\ \phantom{$^5$}Michigan State University, East Lansing, MI 48824-1321, USA\\
$^{3}$International Centre for Radio Astronomy Research -- Curtin University, GPO Box U1987, Perth, WA 6845, Australia\\
$^{4}$Australian Research Council Centre of Excellence for Gravitational Wave Discovery (OzGrav), Clayton, VIC 3800, Australia\\
$^{5}$Center of Excellence for Astrophysics in Three Dimensions (ASTRO-3D), Australia\\
$^{6}$Anton Pannekoek Institute for Astronomy, University of Amsterdam, Postbus 94249, 1090 GE Amsterdam, The Netherlands\\
$^{7}$Instituto de Astronomía, Universidad Nacional Autónoma de México, Ciudad de México, CDMX 04510, Mexico\\
}

\date{Accepted 2021 June 7. Received 2021 April 27; in original form 2020 October 6}

\pubyear{2021}

\begin{document}
\label{firstpage}
\pagerange{\pageref{firstpage}--\pageref{lastpage}}
\maketitle

\begin{abstract}
Hotspots on the surface of accreting neutron stars have been directly observed via pulsations in the lightcurves of X-ray pulsars.  They are thought to occur due to magnetic channelling of the accreted fuel to the neutron star magnetic poles. Some X-ray pulsars exhibit burst oscillations during Type I thermonuclear X-ray bursts which are thought to be caused by asymmetries in the burning.  In rapidly rotating neutron stars, it has been shown that the lower gravity at the equator can lead to preferential ignition of X-ray bursts at this location.  These models, however, do not include the effect of accretion hotspots at the the neutron star surface.  There are two accreting neutron star sources in which burst oscillations have been observed to track exactly the neutron star spin period.  We analyse whether this could be due to the X-ray bursts igniting at the magnetic pole of the neutron star, because of heating in the accreted layers under the hotspot causing ignition conditions to be reached earlier.  We investigate heat transport in the accreted layers using a 2D model and study the prevalence of heating down to the ignition depth of X-ray bursts for different hotspot temperatures and sizes.  We perform calculations for accretion at the pole and at the equator, and infer that ignition could occur away from the equator at the magnetic pole for hotspots with temperature $T_{\mathrm{HS}}\gtrsim1\times10^8\,\mathrm{K}$. However, current observations have not identified such high temperatures in AXPs.

\end{abstract}




\section{Introduction}

Accretion-powered X-ray pulsars (AXPs) are neutron stars in close binary orbits ($P\lesssim1$\,d) that accrete from their companion via Roche-Lobe overflow \citep[e.g.,][]{White1983,Nagase1989,Chakrabarty2005}. In some cases the accretion flow from the accretion disk may be channelled by the magnetic field of the neutron star to the magnetic poles, causing hotspots to form on the surface. These hotspots give rise to X-ray pulsations that are observed in the outburst lightcurves of these systems \citep[e.g.,][]{Wijnands1998}. Detailed timing studies of the pulsations in AXPs provide insight into the neutron star spin period, spin evolution, the binary orbit, and the dense equation of state through mass constraints \citep[e.g.,][]{Chakrabarty1998}. Some AXPs rotate very rapidly, at millisecond spin periods ($\nu\gtrsim100$\,Hz), and thus are known as accretion-powered millisecond pulsars (AMXPs) \citep[see][for a review]{Patruno2012}. AMXPs are pulsars that have been spun up through the process of accretion over time \citep{Tauris2006}. 

Currently, there are 20 known AMXPs that exhibit persistent pulsations in their X-ray lightcurves during outburst, and an additional $\sim$10 AXPs with spin periods between $0.1$--$11$\,Hz \citep[for a review see][]{Patruno2012}. Observationally, AXPs all exhibit relatively faint outburst luminosities, indicating they accrete at low rates ($\dot{m}\lesssim0.1$; \medd\, is the Eddington accretion rate, given by $\medd\approx1.75\times10^{-8}$\,\msun\,yr$^{-1}$ for hydrogen-accreting sources), small donor companion stars ($M_\mathrm{c}\lesssim0.2$\,\msun), and short orbital periods ($P\lesssim1$\,d).  They are all transient systems and accrete into an accretion disk around the pulsar for long periods ($\sim$years).  AXPs exhibit sudden outbursts that last $\sim$1\,month, during which time they become active in X-rays as material is transferred from the disk to the neutron star.

In some AXPs, when the accreted fuel builds up on the surface of the neutron star during outburst, it can ignite unstably, producing a bright ($L\sim10^{38}$\,erg\,s$^{-1}$) thermonuclear explosion known as a Type I X-ray burst \citep[for a review see][]{Galloway2017}.  The conditions that produce an X-ray burst in the accreted layer depend primarily on the temperature, composition of the fuel, and the accretion rate. These basic conditions can be simulated in simple one-dimensional analytic calculations of the accretion column \citep[e.g.,][]{Bildsten1998,Cumming2000}.

Some AXPs exhibit thermonuclear burst oscillations, which are strong oscillations observed in the lightcurve of an X-ray burst \citep[e.g.,][]{Watts2012}. The first observation of this phenomenon was in 1996 in bursts from 4U 1728--34, where a strong 363\,Hz signal was observed in 6 X-ray bursts \citep{Strohmayer1996}. There was an upwards drift in the frequency of these oscillations, and they appeared to disappear near the peak of the burst.  \citet{Strohmayer1996} concluded that these oscillations could be caused by rotational modulation of a bright spot on the surface of the neutron star. We have now observed burst oscillations in many sources and the prevalent theory is that they are caused by highly asymmetric bright patches in the surface layers of a burning neutron star \citep{Watts2012}.  Most burst oscillations observed exhibit a frequency drift during the burst, which could be explained by the brighter-patches location moving over the star (for examples due to travelling waves/modes in the ocean).  There are 2 known sources (XTE 1814--338 and  IGR J17480--2446), however, in which the burst oscillations during the burst rise are phase locked with the accretion-powered pulsations \citep{Watts2008,Cavecchi2011,Motta2011}.

Under the assumption that the observed hotspots are caused by channelled accretion onto the surface of the neutron star at the magnetic pole, the phase locking of burst oscillations with accretion-powered pulsations could be explained by burst ignition occurring at the magnetic pole, and burning being confined to this location during the phase-locking.  This is contrary to the assumption that ignition should preferentially occur at the equator due to the effect of fast rotation on reducing the effective gravity at the equator \citep{Spitkovsky2002,Cooper2007}. 
 
It has been demonstrated that the magnetic fields of XTE 1814--338 and  IGR J17480--2446 are most likely not strong enough to confine material at the magnetic pole, with magnetic confinement requiring $B \gtrsim 4\times10^9 - 3\times10^{10}$\,G \citep{Brown1998, Cavecchi2011} and the magnetic fields of XTE 1814--338 and  IGR J17480--2446 having been estimated to be $0.16\times10^8 - 7.8\times10^{8}$ and $\sim2\times10^8 - 2.4\times10^{10}$\,G, respectively \citep[e.g.,][]{Mukherjee2015,Papitto2012}. Even taking into account dynamical strengthening of the magnetic field \citep{Heng2009}, confinement could still be difficult due to the occurrence of instabilities like the "ballooning" instability \citep[e.g.][]{Litwin2001}. What has not been explored, however, is whether the presence of a hotspot at the surface could induce heating down to the ignition depth of a burst, even after the magnetic confinement has washed out, first suggested as a possibility by \citet{Watts2008}. Furthermore, in models of the ignition location of bursts on the surface of accreting neutron stars, the presence of a hotspot in AXPs is often neglected \citep{Spitkovsky2002,Cooper2007,Cavecchi2017}. 

\begin{figure*}
\includegraphics[viewport=103 15 382 326,clip, width=0.75\columnwidth]{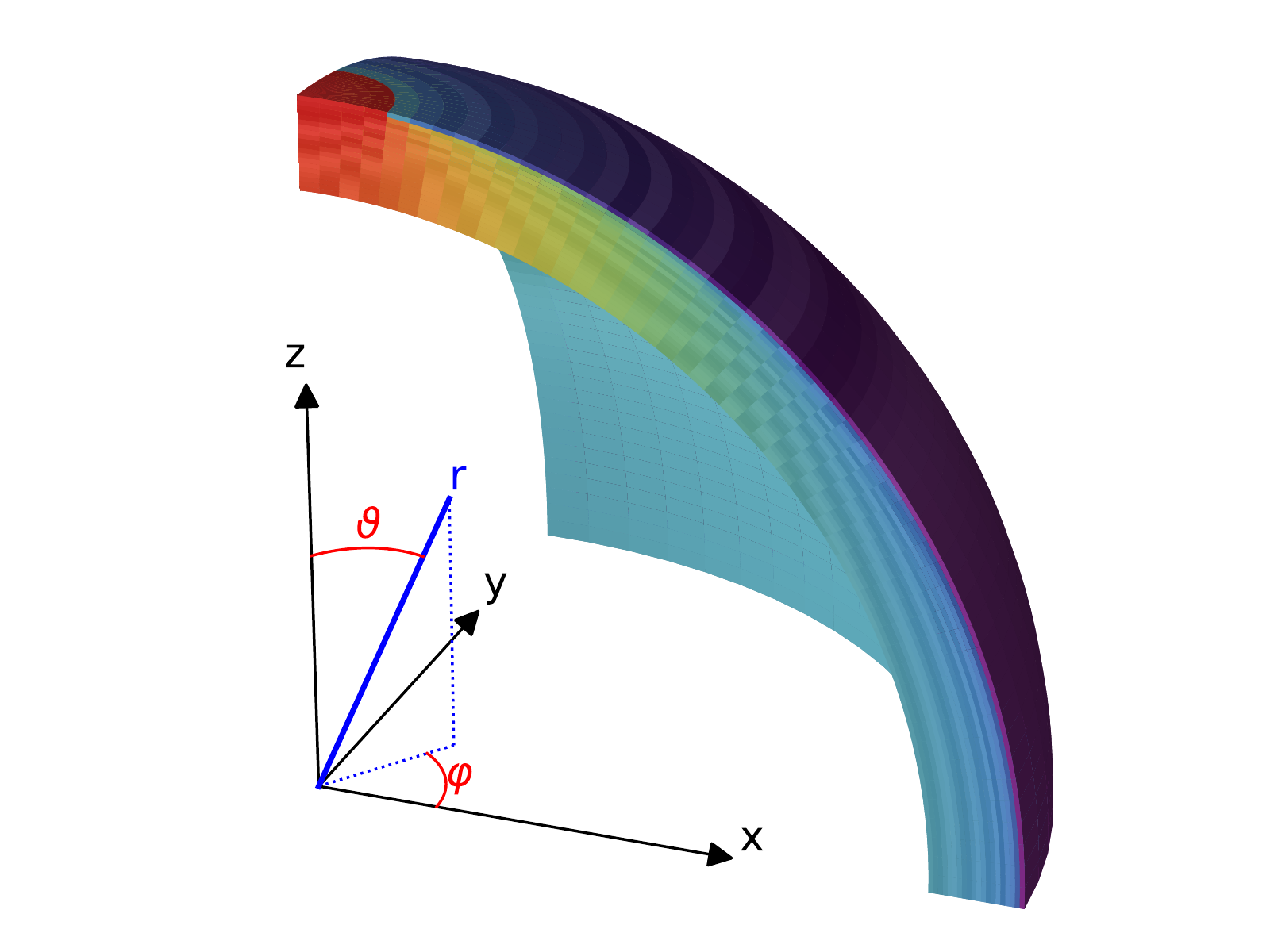}
\hfill
	\includegraphics[viewport=71 51 309 326,clip, width=0.75\columnwidth]{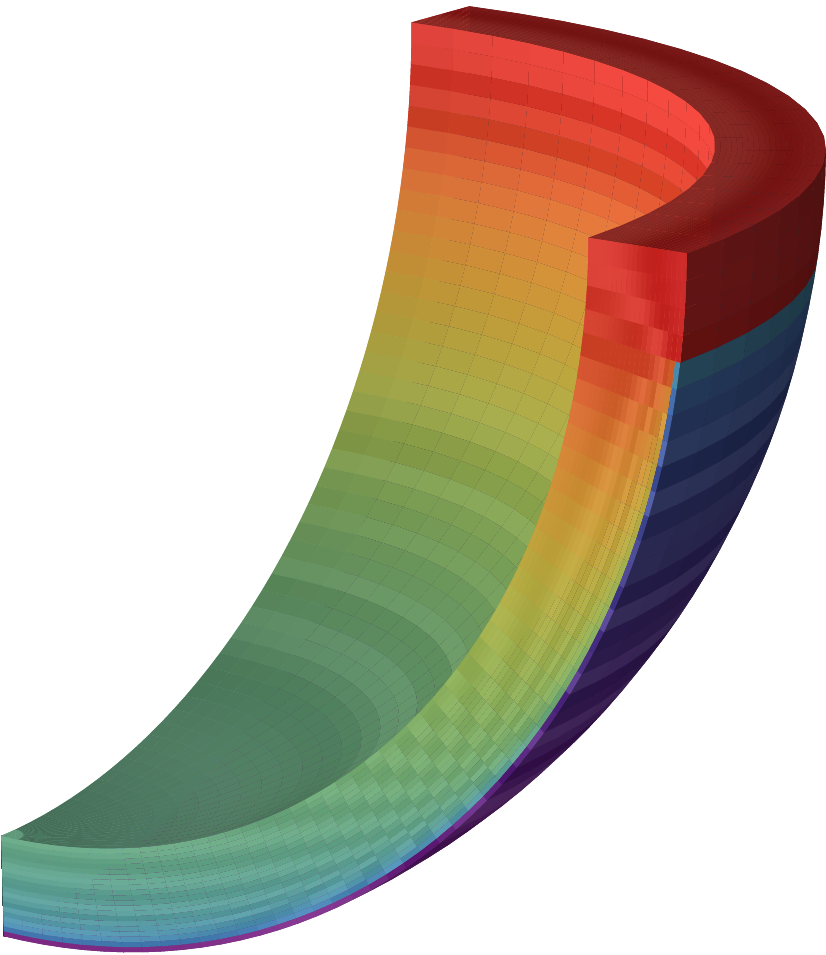}
    \caption{Geometrical setup of polar (\textsl{left}) and equatorial (\textsl{right}) accretion.  These figures are for a simplified setup using a 3D version of our code.  These models are to visualise the geometry only and are not to scale (using $10\,\mathrm{m}$ neutron star radius) nor covering full depths range (outer $1.5\,\mathrm{m}$ shown).  The grid size is $50\times20\times20$ in $\vartheta\times\varphi\times r$.  In both cases the width of the "hot" region (distance from pole or equator, respectively) is $2\,\%$ of the circumference.  Shown is one octant; assume rotational symmetry around the $z$-axis and mirror symmetry about the equator.  Colour coding indicates temperature with the \textsl{red colour} at the hot spot corresponding to $100\,\mathrm{MK}$ (8.6\,keV) and the coolest dark purple colour at the unheated surface having about $0.6\,\mathrm{MK}$ (5.1\,keV).  The base heat flux is $10^{19}\,\mathrm{erg}\,\mathrm{s}^{-1}\,\mathrm{cm}^{-2}$.}
    \label{fig:3D}
\end{figure*}

In this first paper we investigate the plausibility of heating caused by a hotspot or accretion stripe (Figure~\ref{fig:3D}) reaching down to the ignition depth of an X-ray burst, focussing on the case of hydrogen-free accretion as the effect of external heating is likely the most prominent there.  We analyse whether this could cause bursts to preferentially ignite under the hotspot in the accreted layers.  We model the surface layers by solving a 2D heat diffusion equation and exploring the heat transport  inside the accretion column, to determine at what hotspot temperature X-ray bursts would preferentially ignite under the hotspot. 
In Section \ref{sec:methods} we describe the numerical set up and equations used to describe the surface layers of the neutron star, in Section \ref{sec:results} we present the results for different hotspot temperatures and geometries, in Section \ref{sec:discussion} we discuss the limitations of this study, and in Section \ref{sec:conclusion} we provide concluding remarks.

\section{Methods}\label{sec:methods}

We solve the heat diffusion equation in 2D to describe the heat conduction and transport mechanisms in the gaseous accreted layers on the surface of the neutron star.  To derive the diffusive flux we begin with the general diffusive flux for a photon

\begin{equation}\label{eq:dflux}
    j = - D \,\nabla n
\end{equation}
where $D$ is the diffusion coefficient and $n$ is the number density. 

We estimate the diffusion coefficient from a random walk approximation

\begin{equation}\label{eq:Dcoeff}
D = \frac{1}{3}\,v\,l_{\rm{ph}}
\end{equation}
where the velocity, $v$ for a photon may be replaced by the speed of light, $c$, and the mean free path of a photon is $l_{\rm{ph}} = 1/(\rho\kappa)$, where $\rho$ is the density and $\kappa$ is the opacity of the gas. 

$\nabla n$ can be calculated using the energy density of a photon gas, $U = aT^4$, where $a$ is the radiation constant and $T$ is the temperature. Differentiating gives:

\begin{equation}\label{eq:energy}
    \nabla U = a\, \nabla T^4 
\end{equation}
Thus the total diffuse flux, $F$ is given by
\begin{equation}\label{eq:flux}
    F =  \frac{a\,c}{3\,\kappa\,\rho}\,\nabla T^4
\end{equation}
The heat diffusion equation for this flux is thus
\begin{equation}
\label{eq:heatdiffusion}
    \nabla \left(\frac{a\,c}{3\,\kappa\,\rho} \,\nabla T^4\right) = \epsilon(T,\rho)\,\rho
\end{equation}
where $\epsilon(T,\rho)$ is the heat source term (specific energy generation rate), which in this case corresponds to the heating due to nuclear burning.  We assume the simplest case for nuclear burning in accreting neutron stars, in which the accreted fuel is almost pure $^4$He and $\epsilon(T,\rho)$ is given by \citep{Bildsten1998}
\begin{equation}\label{eq:eps}
    \epsilon_{3\alpha} = 5.3\times10^{21}\, \frac{\rho_5^2\,Y^3} {T_8^3} \exp\left(-\frac{44}{T_8}\right)
\end{equation}
where $T_8$ is the temperature in $10^8\,\mathrm{K}$, $\rho_5$ is the density in $10^5\,\mathrm{g\,cm^{-3}}$, and $Y$ is helium mass fraction.  For the accreted material we assume a helium mass fraction of $Y=0.99$ and metallicity of $Z=0.01$ and that the metals do to not contribute to the nuclear energy generation.  There is no nuclear burning in the substrate below the accreted layers. 

We choose to use the triple-$\alpha$ energy generation rate to represent the heating due to nuclear burning in our model since it has been shown that thermally unstable helium burning is likely responsible for the thermonuclear runaway that begins an X-ray burst \citep[e.g.,][]{Bildsten1998}.  We note that there would be additional energy due to CNO burning if hydrogen was present, as well as nuclear burning beyond carbon that we do not account for because it is not significant before runaway in pure helium accretors. 

For the opacity and the density we assume $\kappa$ is a function of $T$ and $\rho$, and $\rho$ is a function of the radial coordinate, $r$ only. 
We extract a density distribution, opacity grid, and initial temperature distribution from a \textsc{Kepler} model, and pre-compute an opacity table for given $T$ and $\rho$, using the mass fractions $Y=0.99$, $Z=0.01$ ($0.009$ $^{14}$N and $0.001$ $^{56}$Fe) for the accreted layers and $^{56}\mathrm{Fe}=1.0$ for the substrate, where the substrate is a thick layer below the accreted layers consisting of a non-reactive substance ($^{56}$Fe in this case). 

\textsc{Kepler} is a 1D multi-zone model of Type I X-ray bursts that integrates the time-dependent equations of conservation of momentum, mass, and energy, in spherical symmetry \citep[see][]{Woosley2004}. \textsc{Kepler} allows for a general mixture of radiation, ions, and degenerate or relativistic electrons, and implements a complex nuclear reaction network. 
For each set of parameters we extract the temperature and density distributions in a snapshot 1 minute before a burst commences, where we define burst commencement as the first timestep in which convection appears in the accreted layer.  For the \textsc{Kepler} model and for the case study of \XTE{} (Model~A), we set the global accretion rate, $\dot{m}=0.1\,\dot{M}_{\mathrm{Edd}}$, hydrogen mass fraction, $X=0.0$, helium mass fraction, $Y=0.99$, metallicity $Z=0.01$ (0.009 as $^{14}$N and 0.001 as $^{56}$Fe), and the base heating $Q_\mathrm{b}=0.1\,\mathrm{MeV}\,\mathrm{nucleon}^{-1}$ and $0.15\,\mathrm{MeV}\,\mathrm{nucleon}^{-1}$. We note that in our simulations we set the global accretion rate across the simulation domain to be constant, at the value expected from the X-ray luminosity of the system.  We then place a hot region at the top to mimic the additional effect of a local higher accretion rate (and heating from the accretion shock above).  An example opacity distribution for the accreted layers and substrate is shown in Figure~\ref{fig:initial}. For the case study of \IGR{} (Model~B), we use a similar \textsc{Kepler} model but set $\dot{m}=0.038\,\dot{M}_{\mathrm{Edd}}$. Here we infer the accretion rate of the system from observations of the X-ray luminosity, $L_\mathrm{X}$ using the relation from \citet{Galloway2008}
\begin{equation}
\label{eq:Mdot}
    \dot{m} = \frac{L_X (1+z)}{4 \pi R^2 Q_{\mathrm{grav}}}
\end{equation}
where $z$ is the gravitational redshift (which we assume is 0.259 for a $1.4\,\mathrm{M}_{\odot}$ neutron star), $R$ is the radius of the neutron star (which we assume is $11.2\,\mathrm{km}$), and $Q_{\mathrm{grav}}$ is the energy released per nucleon during accretion, for which we approximate $Q_{\mathrm{grav}} = GM/R$. 

To model a hot-stripe at the equator we use a \textsc{Kepler} model with the same parameters as the magnetic pole but set the gravity to be $75\,\%$, to emulate the effect of the fast rotation of the neutron star on the effective surface gravity (Model~C). \citet{Spitkovsky2003} suggest that the surface gravity at the equator could be up to $25\,\%$ less than at the poles.

\begin{figure*}
\includegraphics[width=\columnwidth]{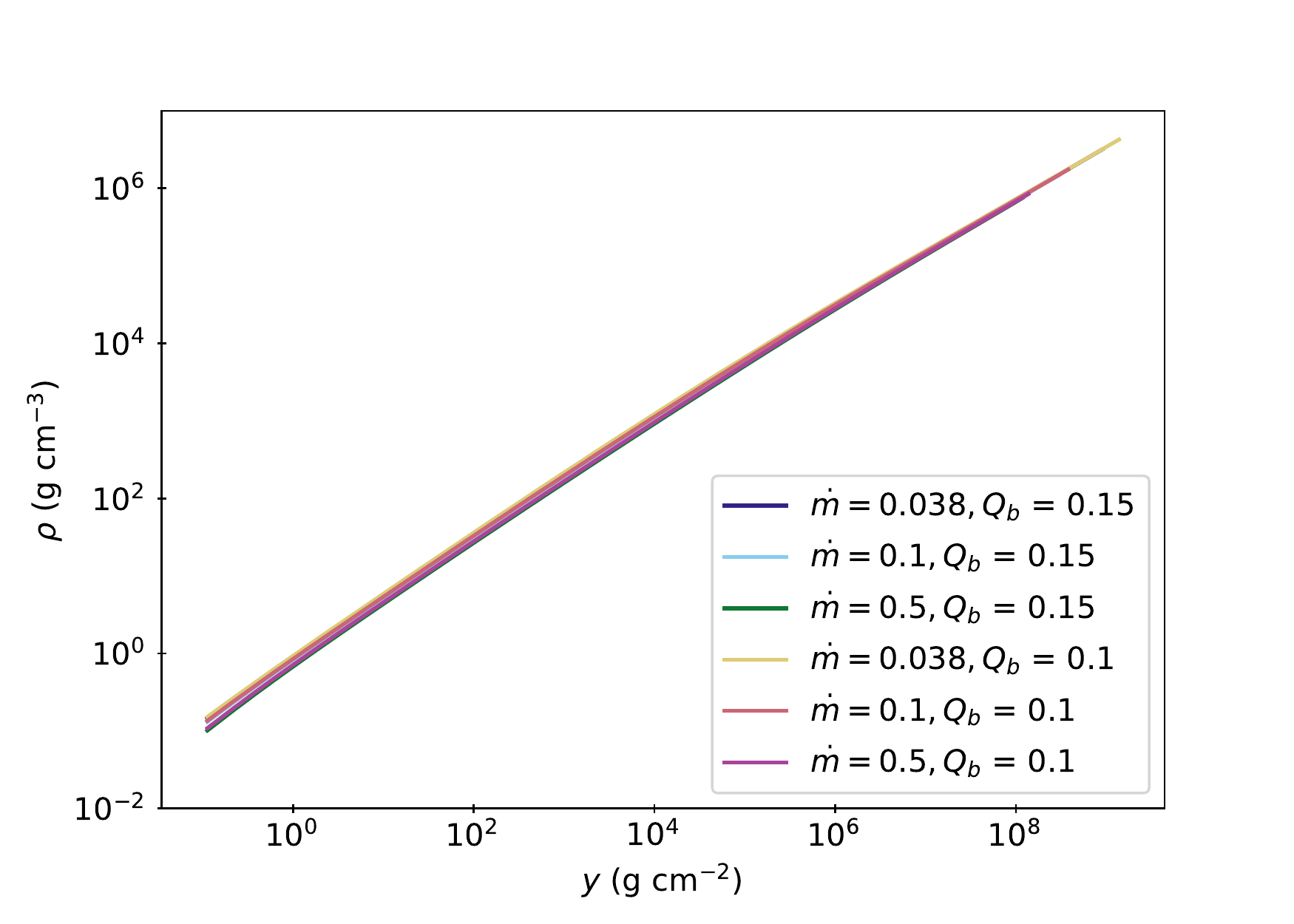}
	\includegraphics[width=\columnwidth]{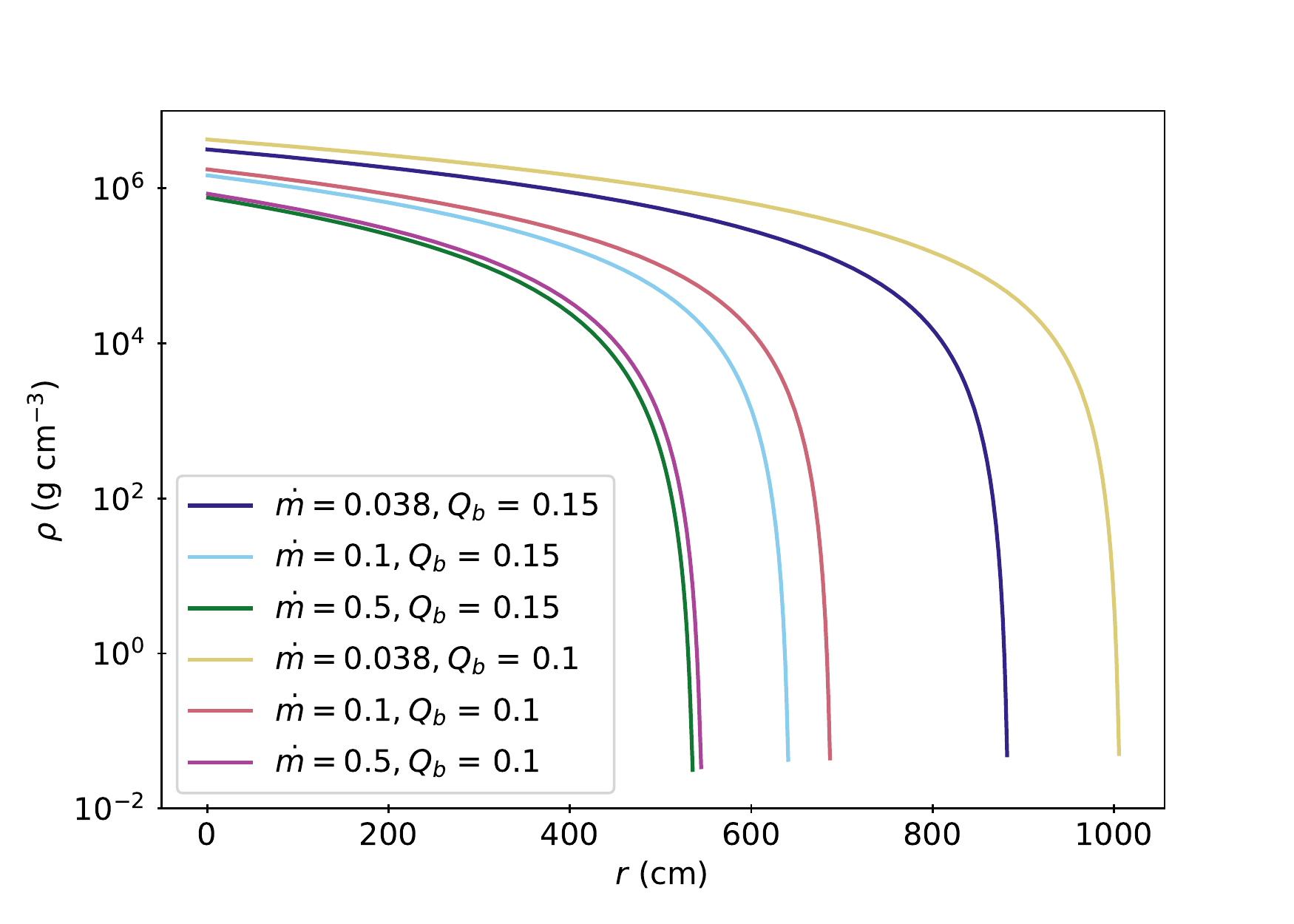}
	\includegraphics[width=\columnwidth]{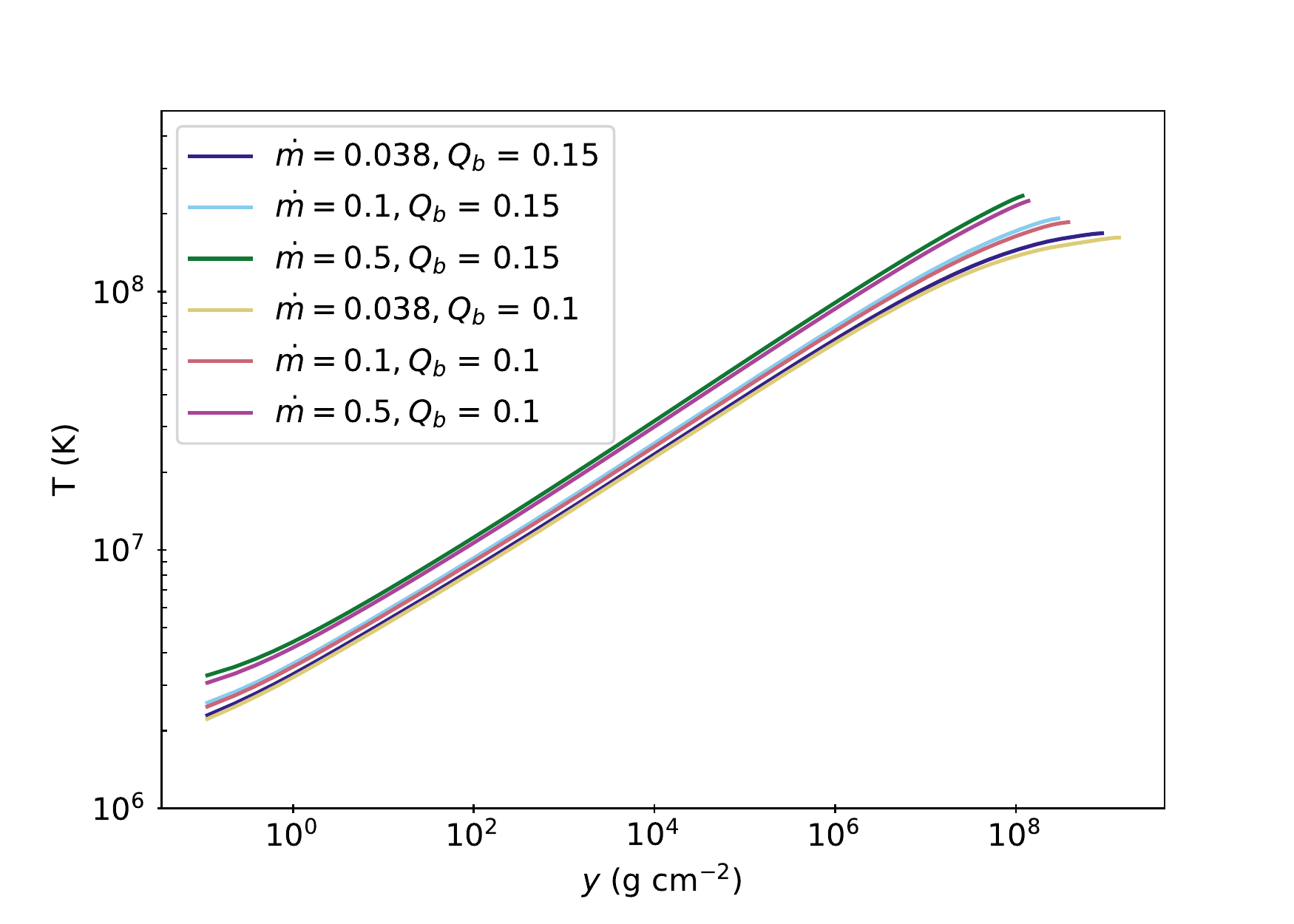}
	\includegraphics[width=\columnwidth]{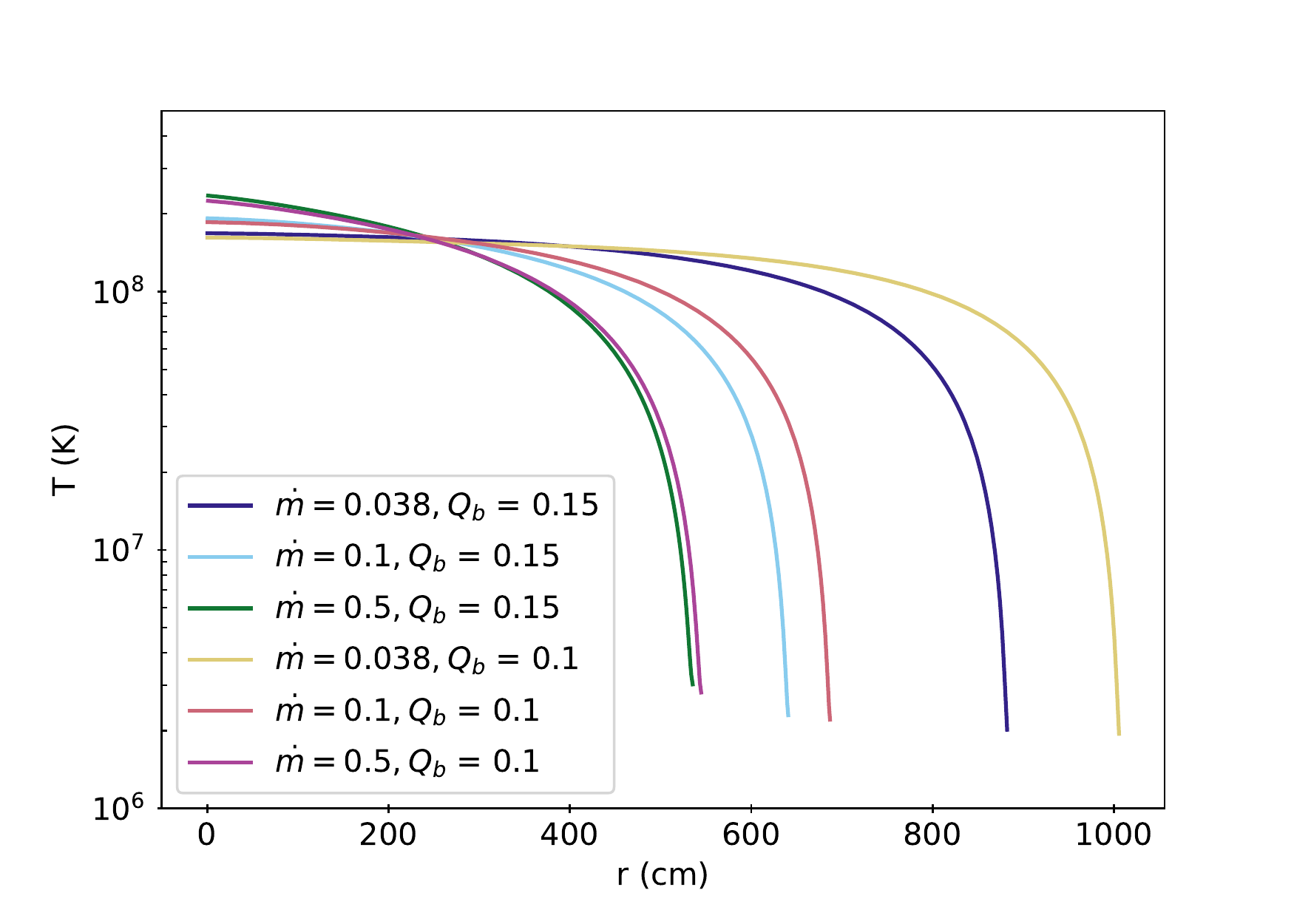}
    \caption{ \textit{Top Panels:}\textsc{Kepler} model distributions of density ($\rho$) for column depth ($y$) and radial coordinate ($r$) for different initial accretion rates ($\dot{m}$) and base fluxes ($Q_{\mathrm{b}}$).
    \textit{Bottom Panels:} \textsc{Kepler} model distributions of temperature ($T$) for column depth ($y$) and radial coordinate ($r$) for different initial accretion rates ($\dot{m}$) and base fluxes ($Q_{\mathrm{b}}$). All distributions are taken from a snapshot 1\,minute before a burst commences.}
    \label{fig:Keprho}
\end{figure*}

The \textsc{kepler} distributions of $\rho$ and $T$ with radial coordinate, $r$, and column depth, $y$, for the accreted layers are shown in Figure~\ref{fig:Keprho}.  For comparison, we have plotted a range of $\dot{m}$ and $Q_\mathrm{b}$ to demonstrate how these distributions change under the assumed parameters.  When the accretion rate is higher, the accreted layer is thinner and hotter than when the accretion rate is lower.  Likewise, when $Q_\mathrm{b}$ is higher, the entire layer is slightly hotter than when $Q_\mathrm{b}$ is lower.  In our models, we ran all cases for the two values of $Q_\mathrm{b}$ (0.1 and $0.15\,\mathrm{MeV}\,\mathrm{nucleon}^{-1}$) but found that since the effect of increasing $Q_\mathrm{b}$ uniformly increased the temperature over the entire layer, there was no significant difference in the results.

\subsection{2D Transport Code}
We solve Equation \ref{eq:heatdiffusion} using a finite volume method to discretise the equation in \textsc{Python}, and set up a two-dimensional grid. We choose to use a finite volume method over a finite difference method to ensure conservation of energy.  

For simplicity and numerical stability, we solve for $\tau:=T^4$ instead of solving for $T$.
To discretise the equations, we integrate over the finite volume of each cell on the grid and apply Gauss's theorem to convert the left-hand term into a surface integral, such that Eq.~\ref{eq:heatdiffusion} becomes
\begin{equation}
    \int_{\Delta A} - \frac{a\,c}{3} 
    \frac{\nabla\tau}{\kappa\,\rho} \,\mathrm{d}A - \int_{\Delta\!V} \rho\,\epsilon(T,\rho) \,\mathrm{d}V = 0
\end{equation}
where $\Delta A$ is the surface area of the cell and $\Delta\!V$ its volume.

The total energy flow into a cell at coordinates $(x,y,z)$ (or corresponding cylindrical or polar coordinates) with indices $(i,j,k)$ is given by the sum of the flows at each of the cell interfaces
\begin{eqnarray}
    F_{i,j,k} &=& {\cal F}_{i+\h,j,k} - {\cal F}_{i-\h,j,k} + {\cal F}_{i,j+\h,k}  \\\nonumber &-&{\cal F}_{i,j-\h,k} +{\cal F}_{i,j,k+\h} - {\cal F}_{i,j,k-\h}
\end{eqnarray}
where the half indices correspond to the zone interfaces.  The flows are centred using the scheme
\begin{eqnarray}
        &&{\cal F}_{i-\h,j,k} = \\\nonumber
        &&-\frac{a\,c\,\Delta\!A_{i-\h,j,k}}
        {3\,\kappa(\rho_{i-\h,j,k}, \tilde{\tau}_{i-\h,j,k})\,\rho_{i-\h,j,k}} \frac{\tau_{i,j,k} - \tau_{i-1,j,k}}{x^\mathrm{c}_{i,j,k}-x^\mathrm{c}_{i-1,j,k}} 
\end{eqnarray}
where $x^\mathrm{c}_{i,j,k}$ denotes the zone centre of zone $(i,j,k)$, $$\tilde\tau_{i-\h,j,k}=\frac12\left(\tau_{i,j,k}+\tau_{i-1,j,k}\right)$$ and $\Delta A_{i-\h,j,k}$ is the surface area in the negative $i$-direction, and $\rho_{i-\h,j,k}$ is the density at the centre of that surface.  In the calculations presented here, the densities are pre-defined and only depend on $z$ coordinate.

With $$\int_{\Delta\!V}\rho\,\mathrm{d}V=m\;,$$ the mass of the zone, the source term from nuclear burning becomes just $m\,\epsilon(T,\rho)$ and our equation to solve becomes
\begin{eqnarray}
F_{i,j,k} - m_{i,j,k}\,\epsilon(\tau_{i,j,k},\rho_{i,j,k}) =   0\;.
\end{eqnarray}

To model a hotspot on the surface we use a 2D cylindrical coordinate system, such that $\Delta A= 2\pi h R_{\mathrm{in}}$ in the $i-1$ direction, $\Delta A= 2\pi h R_{\mathrm{out}}$ in the $i+1$ direction, and $\Delta A=\pi\,(R_{\mathrm{out}}^2 - R_{\mathrm{in}}^2)$ in the $k$-directions, where $h$ is the height, $h=\Delta r$. To model a hot-stripe along the equator we use a Cartesian coordinate system, such that $\Delta A=\Delta r \,\Delta x$ in the $i$ directions and $\Delta A=\Delta x\, \Delta r$ in the $k$-directions. The reason for the choice of these two coordinates systems is due to the assumed symmetry in modelling in 2D.  For a cylindrical coordinate system, symmetry is assumed about the angle coordinate, going around the neutron star, thus simulating a hotspot, and there is rotational symmetry in $j$-direction (see Figure~\ref{fig:3D}).   Whereas, in a Cartesian coordinate system, translational symmetry is assumed in $j$-direction (i.e., the $y$ direction), and thus heat cannot diffuse in this direction (rather it is constant), which simulates a stripe along the equator, allowing us to just  model a cross-section of this stripe. 

We solve for $\tau$ by relaxation of an initial guess. We define a coefficient matrix, $A$ such that $A \tau = b$, where $b$ is a matrix consisting of the source terms and A is the Jacobian matrix. We use the sparse matrix solver of the SciPy \textsc{Python} package \citep{2020SciPy}.

The grid is uniformly separated in $x$ and non-uniform in $r$, for which we use the zone-width from the \textsc{Kepler} snapshots, which results in higher resolution close to the surface. For all models we use a grid of $x\times r$ = $280\times(220-280)$, where the number of $r$ zones depends on the specific \textsc{Kepler} model parameters.  We ran tests with lower and higher resolution and found no significant changes in the results.

\subsubsection{Boundary Conditions and Initial Conditions}
The proper choice of boundary conditions significantly affects our results.  On the sides perpendicular to $i$ we use reflective boundary conditions.  We simulate the hotspot on top on one side of the box, since the reflective boundary conditions ensure symmetry in the cylindrical radial coordinate.  For the surface boundary we use a blackbody with temperature corresponding to the surface temperature of the \textsc{Kepler} model.  The surface temperature in the model depends on how far into the neutron star atmosphere we wish to model.  Since we extract the density distribution from \textsc{Kepler}, we set the surface at an optical depth of $\nicefrac23$, the photosphere of the star. The surface at this location has a density of $\sim1$\,g\,cm$^{-3}$, and temperature of $\sim10^6$\,K.  Modelling into the photopshere of the neutron star aids in modelling a hotspot due to the accretion shock, rather than a hotspot deep within the neutron star atmosphere.  For the base boundary we use both a constant flux condition, where the base flux is set by our choice of $Q_\mathrm{b}$, for which we use the commonly used value of $0.1\,\mathrm{MeV}\,\mathrm{nucleon}^{-1}$ (but also tested $0.15\,\mathrm{MeV}\,\mathrm{nucleon}^{-1}$ in all cases), or a constant temperature condition. We found no significant difference in the depth of heating induced by the hotspot between models with a constant flux base boundary and a constant temperature base boundary, but we found the models with a constant flux base boundary reached runaway in general a couple of seconds earlier than those with a constant temperature boundary. The reason for such a  little difference between the runs with the two kinds of base boundary conditions is the fact that we simulate a substrate of Iron at the bottom of the domain, which sets the boundary deep, and far enough from the region we are interested in modelling.  We do not model the core of the neutron star, however, we expect some heat to diffuse lower into (or upwards out of) the neutron star than our simulation bounds. Thus, we find a constant flux base boundary is more physically motivated, as this allows heat to move into or out of our simulated domain. For all models presented in the results, we use a constant flux boundary at the bottom of our domain.

\subsection{Physical sizes of heating regions}
The size of the hotspot can be approximated by the size of the polar cap \citep[e.g.,][]{Bogdanov2016}
\begin{equation}
  \label{eq:r-polar-cap}
  R_{\mathrm{pc}}
  = \sqrt{ \frac{2 \pi R^3}{P c}}
\end{equation}
where $R$ is the radius of the star, $P$ is the spin period, and $c$ is the speed of light. 

The hotspot may be smaller than the polar cap.  For IGR J17480--2446, assuming $R=11.2$\,km\footnote{Throughout this work we consistently assume $R_{\mathrm{NS}}=11.2$\,km. We base this value off the ranges predicted by \citet{Steiner2013} for accreting neutron star sources, but note that the actual neutron star radius in each of these systems is likely different \citep[see e.g.,][for a millisecond pulsar radius measurement]{Riley2019}.} and $P=0.0909\,\mathrm{s}$ ($\nu=11\,\mathrm{Hz}$), we find  $R_{\mathrm{HS}}\lesssim R_{\mathrm{pc}}\approx0.57\,\mathrm{km}$.  For XTE J1814--338, assuming $R=11.2$\,km and $P=0.00318\,\mathrm{s}$ ($\nu=314\,\mathrm{Hz}$), we find $R_{\mathrm{HS}}\lesssim R_{\mathrm{pc}}\approx3.04\,\mathrm{km}$.  Thus we set the width of our simulation to $4\,\mathrm{km}$, and the size of the hotspot to be the relevant polar cap size ($R_{\mathrm{pc}}$). We note that it is extremely likely that the hotspot size in these two systems is smaller than these polar cap estimates, especially due to the high fractional amplitude of the accretion-powered pulsations in \XTE. However, in section \ref{sec:hsize} we demonstrate that once the hotspot is larger than $\sim100$\,m, the size of the hotspot no longer affects the base temperature of the stationary solution, so the size of the domain, and the size of the hotspot once it exceeds 100\,m, do not affect the results.

Since the ocean and atmosphere layer depth is of the order $H\sim10\,\mathrm{m}$, there is at least a factor of $100$ difference between the size of the spot and the depth needed to be considered in the ocean.  For the depth of our box, we use the entire accreted layer, extracted from the \textsc{Kepler} simulations of the accretion column up to the point just before runaway occurs, which extends down to a column depth of $y\approx3\times10^8$\,\gcm, as well as the substrate that the accreted material is accreted onto, which extends down to $y\sim10^{12}$\,\gcm.

For the hotspot temperature, we explore a range of temperatures, noting that for the AMXP SAX J1808.4--3658, the hotspot temperature has been inferred to be $0.9\times10^7$\,K \citep{poutanen2003}.  The hotspot in XTE~1814--338 and  IGR~J17480--2446 should theoretically be slightly hotter than SAX~J1808.4--3658 as these sources have slightly stronger magnetic fields and so could more strongly channel the accretion, producing a hotter hotspot.  The temperature of the hotspot, however, also depends on the size of the hotspot, and the accretion rate.  Spectral modelling of the persistent X-ray emission from IGR J17480--2446 in outburst has estimated temperatures of $0.7-0.9$\,keV \citep{Papitto2012}.  For XTE~J1814--338, \citet{Krauss2005} measured a non-burst hotspot temperature of $0.96\pm0.13$\,keV, whereas \citet{Bhattacharyya2005} find that a hotspot temperature of $\sim2.3\times10^7$\,K is reasonable for this system during a burst, and spectral modelling has estimated the X-ray emission to have a temperature of $2-3.5\times10^7$\,K \citep{Strohmayer2003}. We note that the hotspot temperature measurement most relevant to our models is that measured outside of bursts. Here we explore a range of hotspot temperatures from  $2\times10^7\,\mathrm{K}$ to  $1.2\times10^8\,\mathrm{K}$.  

\begin{figure}
    \includegraphics[width=\columnwidth]{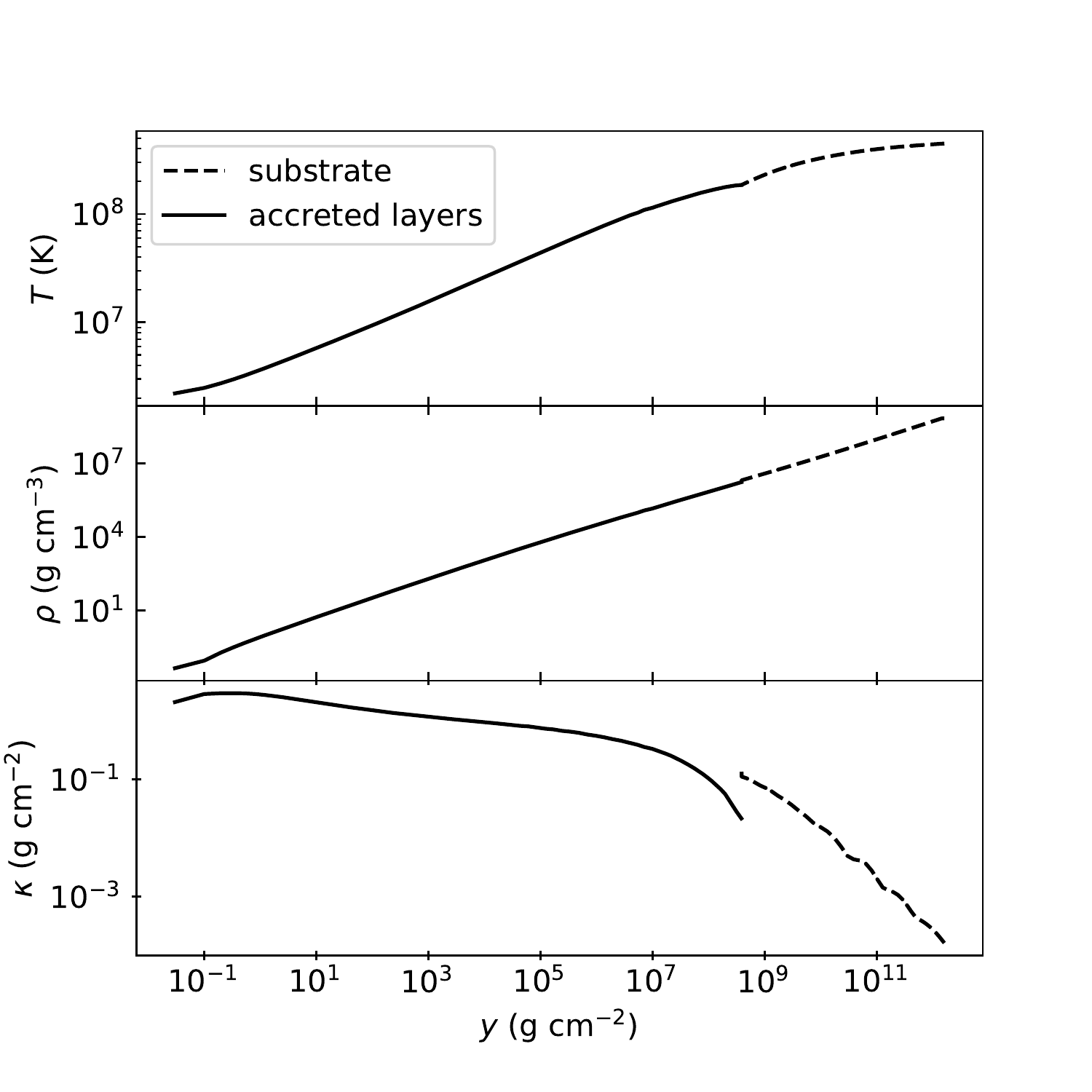}

    \caption{Initial distribution of temperature (top), density (middle) and opacity (bottom) with column depth $y$. The temperature and density distributions are extracted directly from a \textsc{kepler} model with parameters  $\dot{m}=0.1$\,$\dot{M}_{\mathrm{Edd}}$, $Q_\mathrm{b}=0.1$, $Y=0.99$, and $Z=0.01$, 1 minute before a burst begins. The opacity is extracted from our opacity grid for given $T$ and $\rho$. The dashed line indicates the substrate and the solid line indicates the accreted layers. The jump in opacity curve at the substrate boundary is due to discontinuous change in composition from He to Fe.}
    \label{fig:initial}
\end{figure}

We initialise the model with the temperature distribution from the relevant \textsc{Kepler} model, shown in Figure~\ref{fig:initial}, except the cells along the top boundary where the hotspot is placed, and solve for the steady-state solution. In reality, a static solution for temperature does not exist at the time of a burst, due to the nature of thermonuclear runaway. However, including time-dependence and modelling a full runaway is outside the scope of this work and so we solve for a static solution as close to runaway as possible. Thus we are able to examine the depth to which heating persists during accretion just prior to a burst, and operate under the assumption that the conditions just prior to a burst will not be significantly different to the conditions at the time of a burst. 

\subsection{Ignition Depth}
We use three different approaches to independently verify the ignition properties of the model calculations. First, to infer the ignition depth via temperature of a burst, we assume the radial coordinate, $r$, is related to the column depth, $y$, by $\mathrm{d}y = \rho\,\mathrm{d}r$ \citep[e.g.,][]{Bildsten1998,Cumming2000}.  One may estimate the temperature of the ignition depth using the simple model of \citet{Bildsten1998} for pure helium burning:
\begin{equation}
\label{eq:ignition}
    T_{\mathrm{ign}} = 1.83 \times 10^8\,\mathrm{K}\cdot \kappa_0^{-\nicefrac1{10}}\, Y_{\phantom0}^{-\nicefrac3{10}} \mu_0^{-\nicefrac15}\, g_{14}^{-\nicefrac15}\, y_8^{-\nicefrac25}
\end{equation}
where $g_{14}$ is the surface gravity in $10^{14}\,\mathrm{cm}\,\mathrm{s}^{-2}$, $y_8$ is the column depth in $10^8\,\mathrm{g}\,\mathrm{cm}^{-2}$, $\kappa_0$ is the opacity in $\mathrm{cm}^2\,\mathrm{g}^{-1}$, $Y$ is the helium mass fraction of the accreted material, and $\mu_0$ is the mean molecular weight of the accreting gas in $\mathrm{g}\,\mathrm{mol}^{-1}$.  This model only provides a crude guidance; the \textsc{Kepler} model did not experience runaway for these conditions, and our models find stationary solutions well beyond that point; a stationary solution is only possible when there is no runaway.  

Thus we also independently infer when runaway occurs by calculating the stationary solution using our 2D heat transport code for a sequence of \textsc{Kepler} dumps consecutively closer to the start of the burst (when convection begins in \textsc{Kepler}). When there is no stationary solution, the model should not converge.   We compare the time before the start of convection at which no stationary solution is found for a model with no hotspot to the same time for models with hotspots of varying temperatures. If a model with hotspot fails to find a stationary solution before a model without hotspot fails, this may indicate that heating from the hotspot has induced runaway earlier. Obviously, it is difficult to assess that no solution could be found with a different numerical scheme, so our inferred times form the 2D models have to be considered upper limits on the change in runaway time.  To optimise out search for a stationary solution, we set the initial guess temperature distribution for the hotter hotspots to be the solution for a slightly cooler hotspot, which in some cases allows the code to find a solution.  At some point, however, ignition conditions are met and a static solution is not possible to find with our approach, which is the time we report.  The equations may still have a solution, the fully-activated burning, but that may lie outside the physics domain of our burning physics and tabulated opacity tables.

Finally, we independently infer when runaway occurs under the influence of a hotspot by carrying out modified \textsc{Kepler} calculations of the lead up to a burst. For each set of parameters, we explore a range of hotspot temperatures by setting the surface boundary to the hotspot temperature (since \textsc{Kepler} is 1D it does not model horizontal diffusion as in the 2D heat transport code), we set the surface boundary pressure to be 3$P_{\mathrm{rad}}$, where $P_{\mathrm{rad}}=\sigma T^4$ is the radiation pressure, and lower the surface resolution to allow convergence (see Footnote~\textsl{b}, Table~\ref{tab:ignitiontime}).  We define the commencement of thermonuclear runaway as the time at which convection first begins. All other model parameters are the same as described for the 2D transport code.  As we show later, for sufficiently large (realistic) hotspots, the centre of the hotspot is not affected by heat loss at the edges, and therefore these \textsc{Kepler} simulations are an adequate approximation for the core of the hotspot or stripe.

\section{Results}\label{sec:results}

A summary of the time before ignition in the KEPLER models without hotspot is reached for the three different models explored is shown in Table~\ref{tab:ignitiontime}.

\begin{table}
	\begin{center}
	\caption{
	Earliness of thermonuclear runaway caused by hotspot heating for different accretion rates, geometries, and hotspot temperatures.}
	\label{tab:ignitiontime}
	\begin{tabular}{llrrr} 
	 \hline
		Model & $\dot{m}$ & $T_{\mathrm{HS}}$ & 2D code$^\mathrm{a}$ & \textsc{Kepler}$^\mathrm{b}$ \\
		Case & (\medd) & (MK) & (s) & (s)\\
		\hline
		
		A   & 0.1 & No hotspot & 0 &(-17.7)$^\mathrm{c}$\\
		pole&0.1 & 20  & 0   & 20.0 \\
		    &0.1 & 50  & 0   & 103\\
		    &0.1 & 100  & 0   & 3,009\\
		    &0.1 & 115 & 5   & 5,122\\
		    &0.1 & 120  & $>60$ & 5,949\\
		
		\hline
	    B&0.038 & No hotspot & 0 & (-1.0)$^\mathrm{c}$\\
		pole&0.038 & 20 & 0 & 47.5 \\
		&0.038 & 50 & 0 & 4,554\\
		&0.038 & 100 & 7--8 & $9.31\times10^4$\\
		&0.038 & 115 & 39--58 & $1.30\times10^{5}$ \\
		&0.038 & 120 & $>60$ & $1.41\times10^{5}$\\
		\hline
		C&0.1 &No hotspot & 76--80 & (-27.8)$^\mathrm{c}$\\
		equator&0.1 & 20 & 81--106 & 4.9\\
		($0.75\,g$)&0.1 & 50 & 81--106 & 34.0\\
		&0.1 & 100 & $>107$ & 955\\
		&0.1 & 115 & $>107$ & 1,782 \\
		&0.1 & 120 & $>107$ & 2,101\\
	
		\hline
	\end{tabular}\vspace{-0.5\baselineskip}
	\end{center}
		\small{
		$^\mathrm{a}$ measured as time of \textsc{Kepler} model used in the 2D calculation at which no stationary solution is found relative to the \textsc{Kepler} model in which convection first appears (i.e. $t=0$ is the \textsc{Kepler} model in which convection first appears, and the time given is how much earlier than this runaway occurs in the 2D code).\\
	$^\mathrm{b}$ time of \textsc{Kepler} model with heating when convection sets in relative to the \textsc{Kepler} model in which convection first appears without heating.\\		
	$^\mathrm{c}$ due to adjusted boundary condition there is a small offset relative to the base model for 2D runs; all other \textsc{Kepler} values in this column are measured relative to this reference.  Total time to burst is $2.527\times10^{4}\,\mathrm{s}$, $2.459\times10^{5}\,\mathrm{s}$, and $1.490\times10^{4}\,\mathrm{s}$ for Models A--C, respectively.}
\end{table}

\subsection{The case of XTE J1814--338}

The observed and inferred system parameters for \XTE\, are shown in Table~\ref{tab:1814obs}.  Based on these observed parameters, we set up our model such that $R_{\mathrm{HS}}=3\,\mathrm{km}$, $\dot{m}=0.1$\medd, we explored a range of hotspot temperatures, and call the models with this set of parameters Model~A.

\begin{table}
	\centering
	\caption{
	Observed and inferred parameters for \XTE}
	\label{tab:1814obs}
	\begin{tabular}{llll} 
		\hline
		Parameter & Value & Units & Ref. \\
		\hline
		$P$& 314 & Hz & [1]\\
		$F_{\mathrm{X}}^{*}$ & $(0.3-0.51)\times10^{-9}$ & ergs\,s$^{-1}$\,cm$^{-2}$ & [1,2] \\
		$d$ & 8 & kpc & [2] \\
		$\dot{m}$& 0.04--0.1 &$\dot{M}_{\mathrm{Edd}}$ & Eq.~\ref{eq:Mdot}  \\ 
		$R_{\mathrm{pc}}$ & 3.04 & km & Eq. \ref{eq:r-polar-cap} \\
		$k_{\mathrm{B}}T$ (burst) & 1.7--3& keV & [2]\\
		$k_{\mathrm{B}}T$ (non-burst) & 0.95$\pm$0.13& keV & [3] \\
		$B$ & $(0.16-7.8)\times10^{8}$ & G & [1], [4] \\
		\hline
	\end{tabular}
	\small{\\$k_\mathrm{B}$ is the Boltzmann constant, $F_{\mathrm{X}}$ is the persistent X-ray flux, $B$ is the magnetic field strength. $^*$This flux range encompasses values measured through the outbursts. {Ref.:} [1] \citet{Papitto2007}, [2] \citet{Strohmayer2003} , [3] \citet{Krauss2005} ,  [4] \citet{Mukherjee2015}
}
\end{table}

The time before convection at which no stationary solution was found for \textsc{Kepler} dumps sequentially closer to runaway is shown in Table~\ref{tab:ignitiontime} (Model~A) for different hotspot temperatures.  For the 2D code we found that only for hotspot temperatures $T_\mathrm{HS}\gtrsim1.15\times10^8\,\mathrm{K}$ did runaway occur earlier than the case with no hotspot, and we found that runaway occured 5 seconds earlier for a hotspot of $1.15\times10^8\,\mathrm{K}$ and more than 60 seconds earlier for a hotspot of $2\times10^8\,\mathrm{K}$.  In the \textsc{Kepler} calculation under the same conditions, however, we found that a hotspot of just $2\times10^7\,\mathrm{K}$ was sufficient to induce runaway 20\,s earlier than the case with no hotspot. Given the timescales that X-ray bursts usually operate on ($\sim60\,\mathrm{s}$), we deduce that an X-ray burst igniting $5$--$20\,\mathrm{s}$ earlier at the hotspot would be sufficient to induce burning located at the magnetic pole.

\begin{figure*}
	\includegraphics[width=\columnwidth]{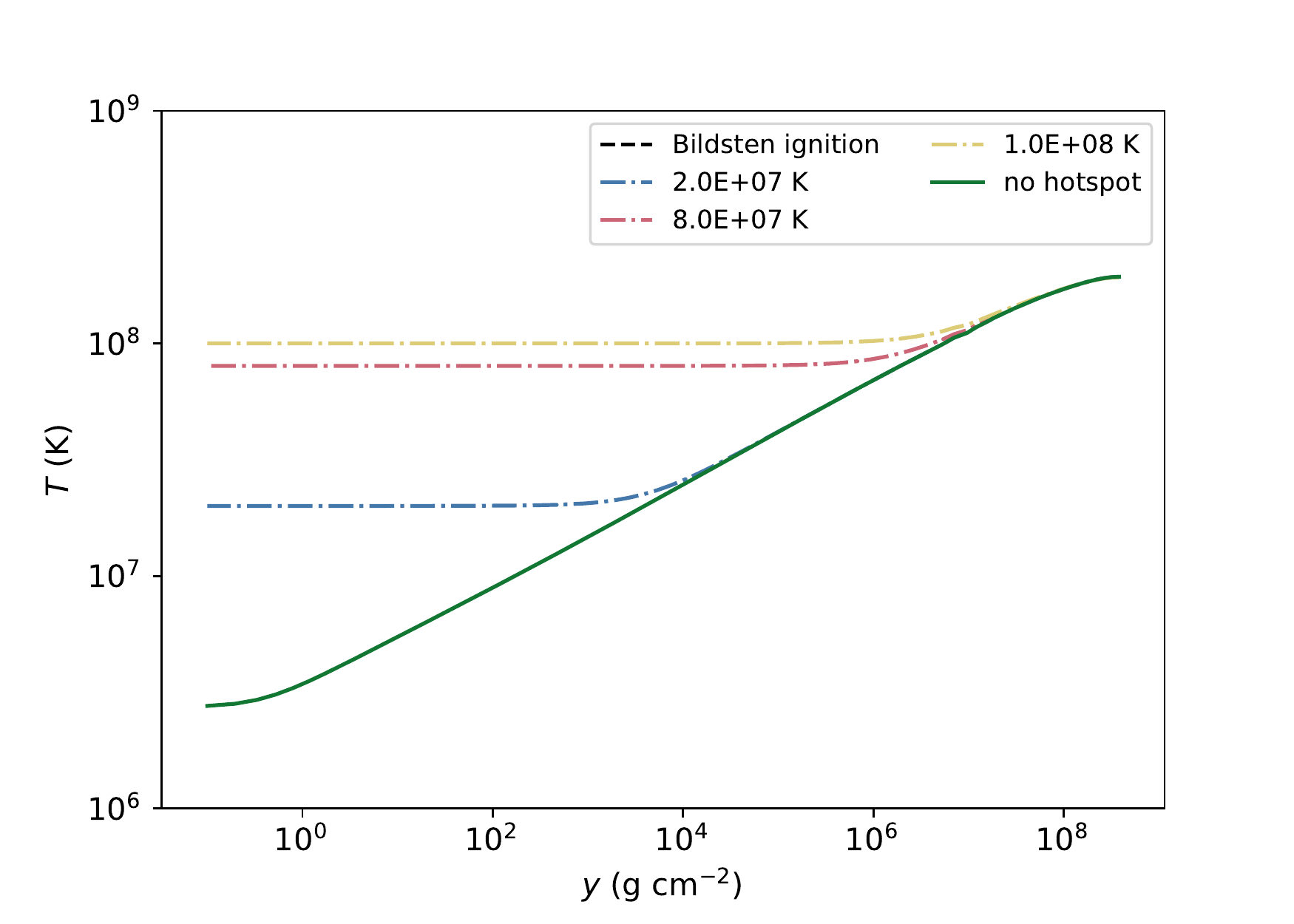}
	\includegraphics[width=\columnwidth]{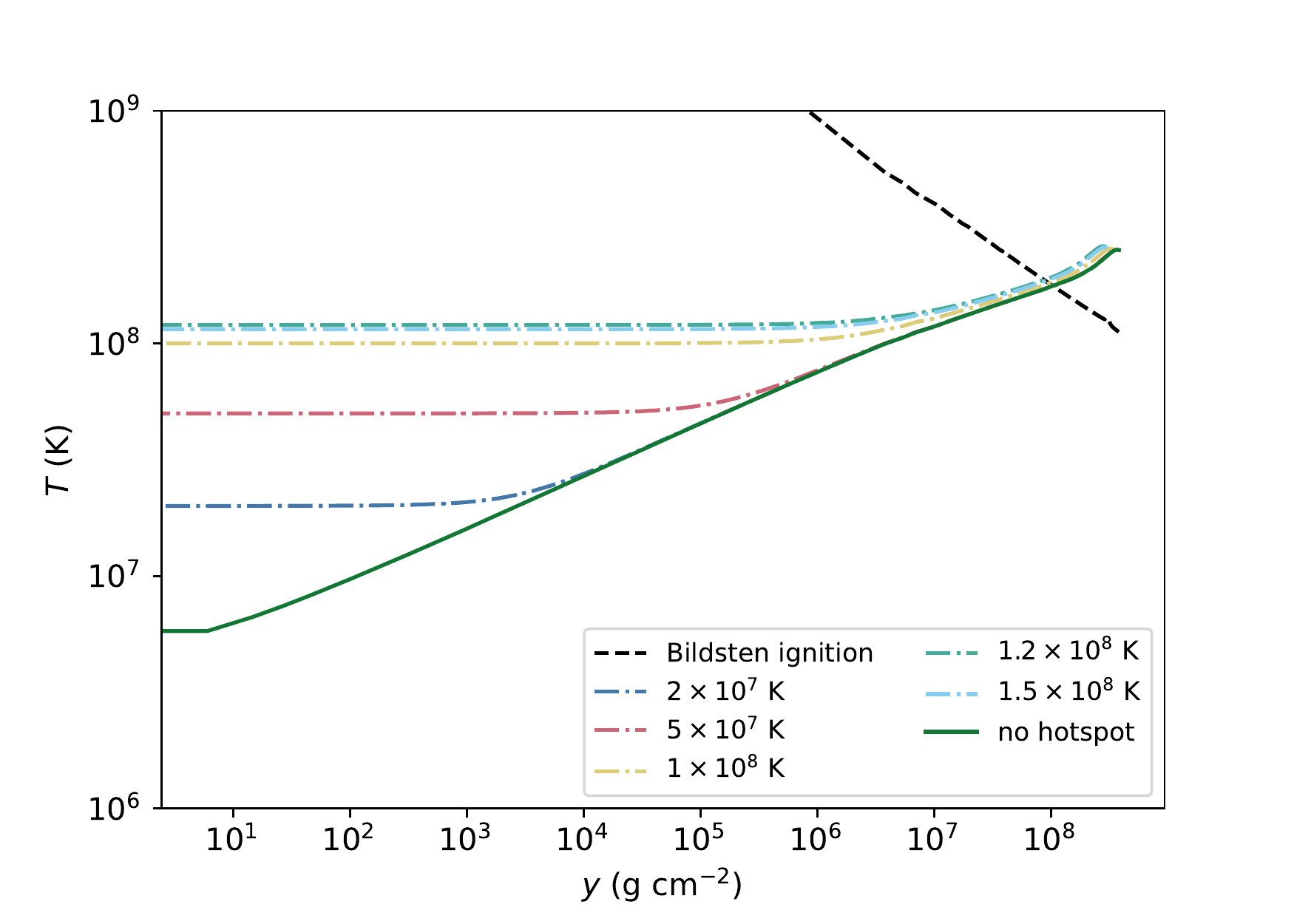}
	\includegraphics[width=\columnwidth]{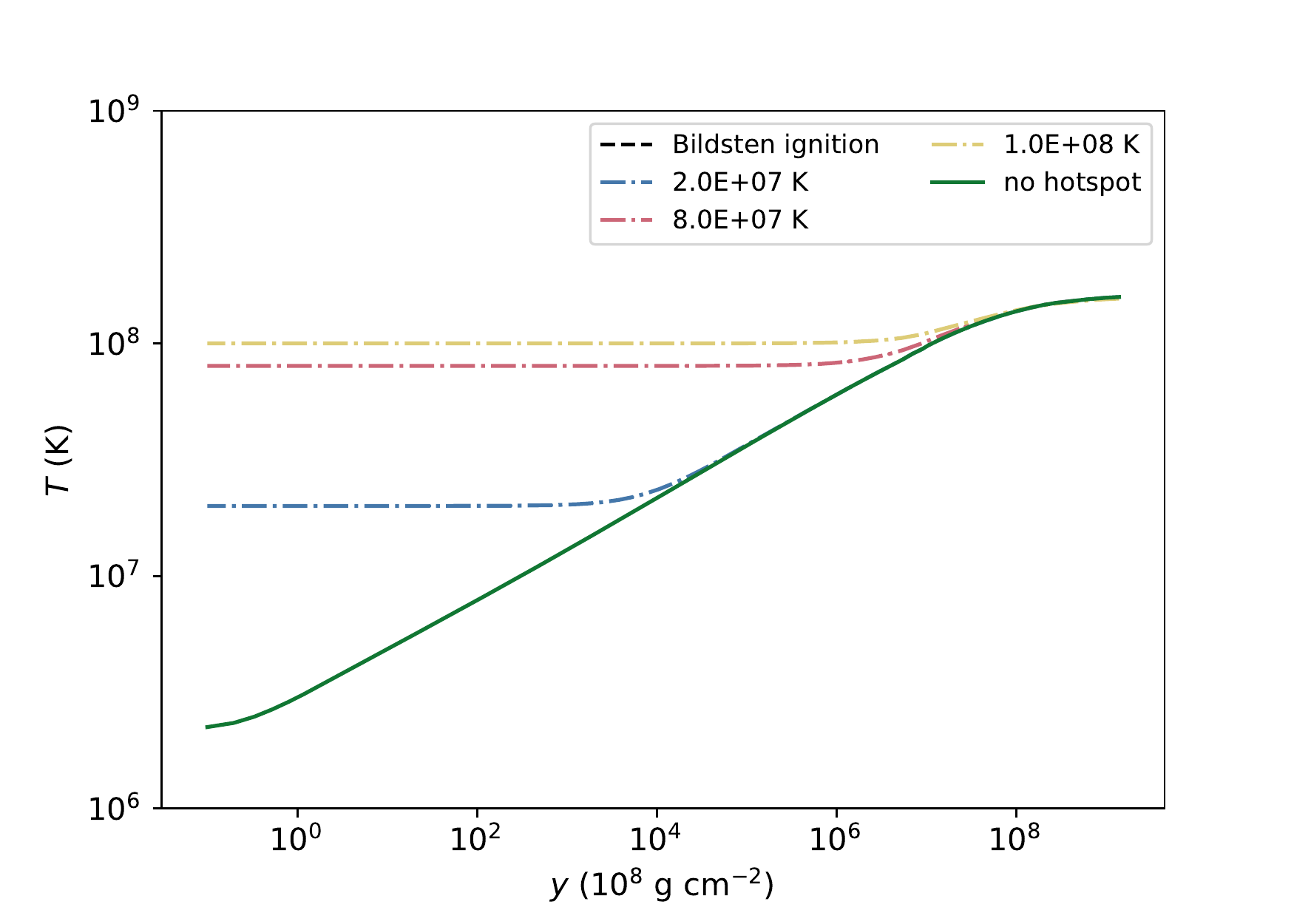}
 	\includegraphics[width=\columnwidth]{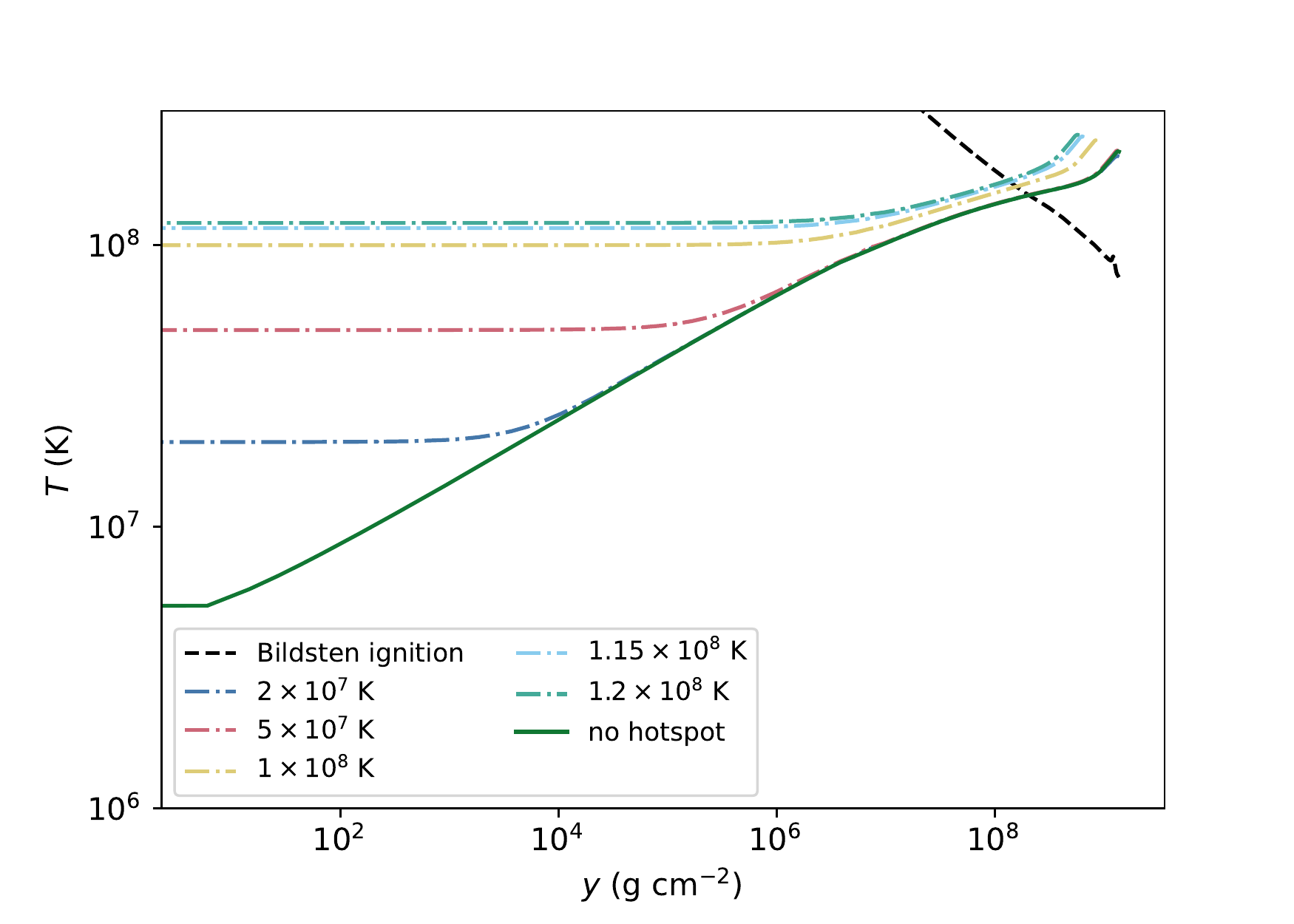}
 	\includegraphics[width=\columnwidth]{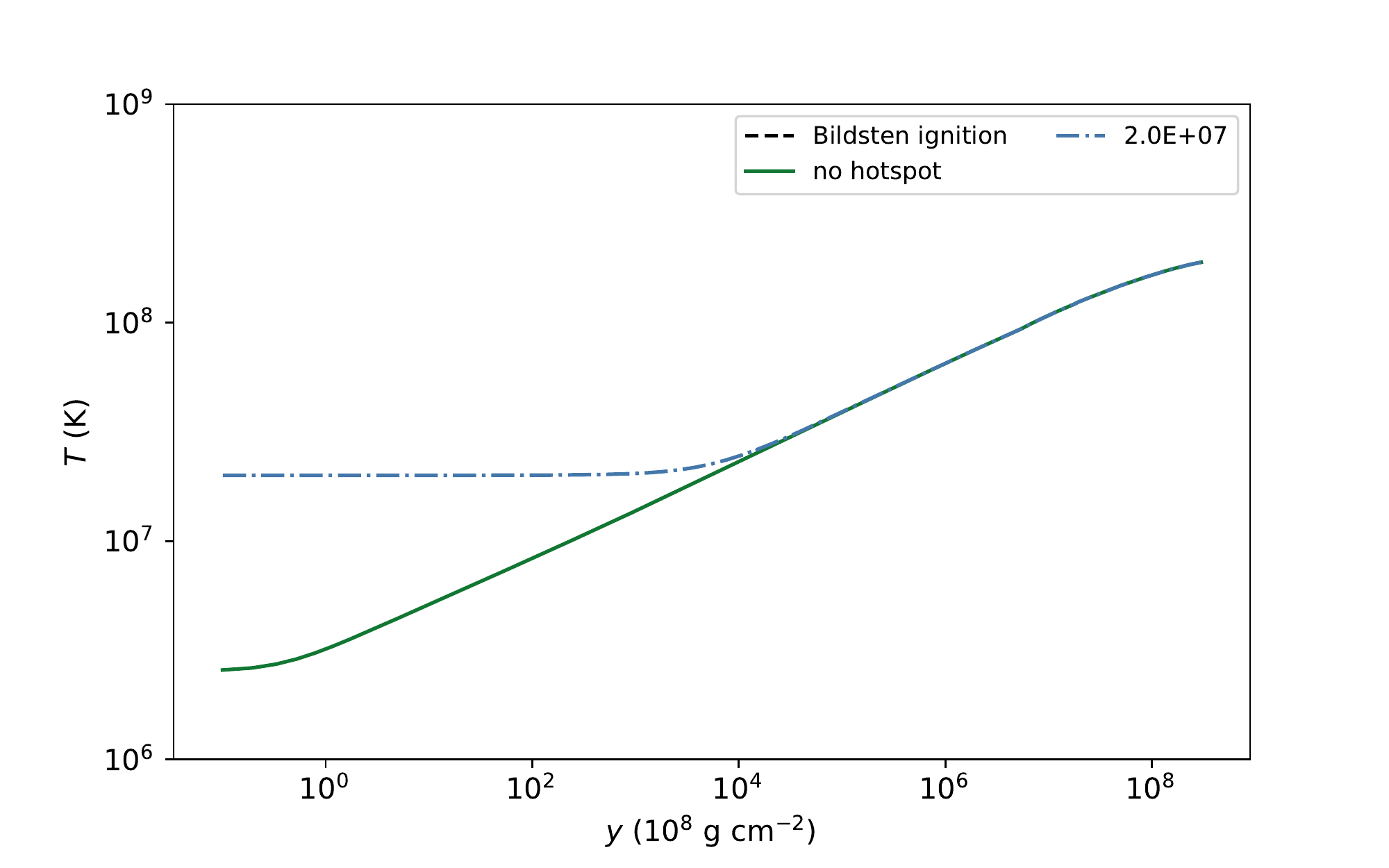}
 	\includegraphics[width=\columnwidth]{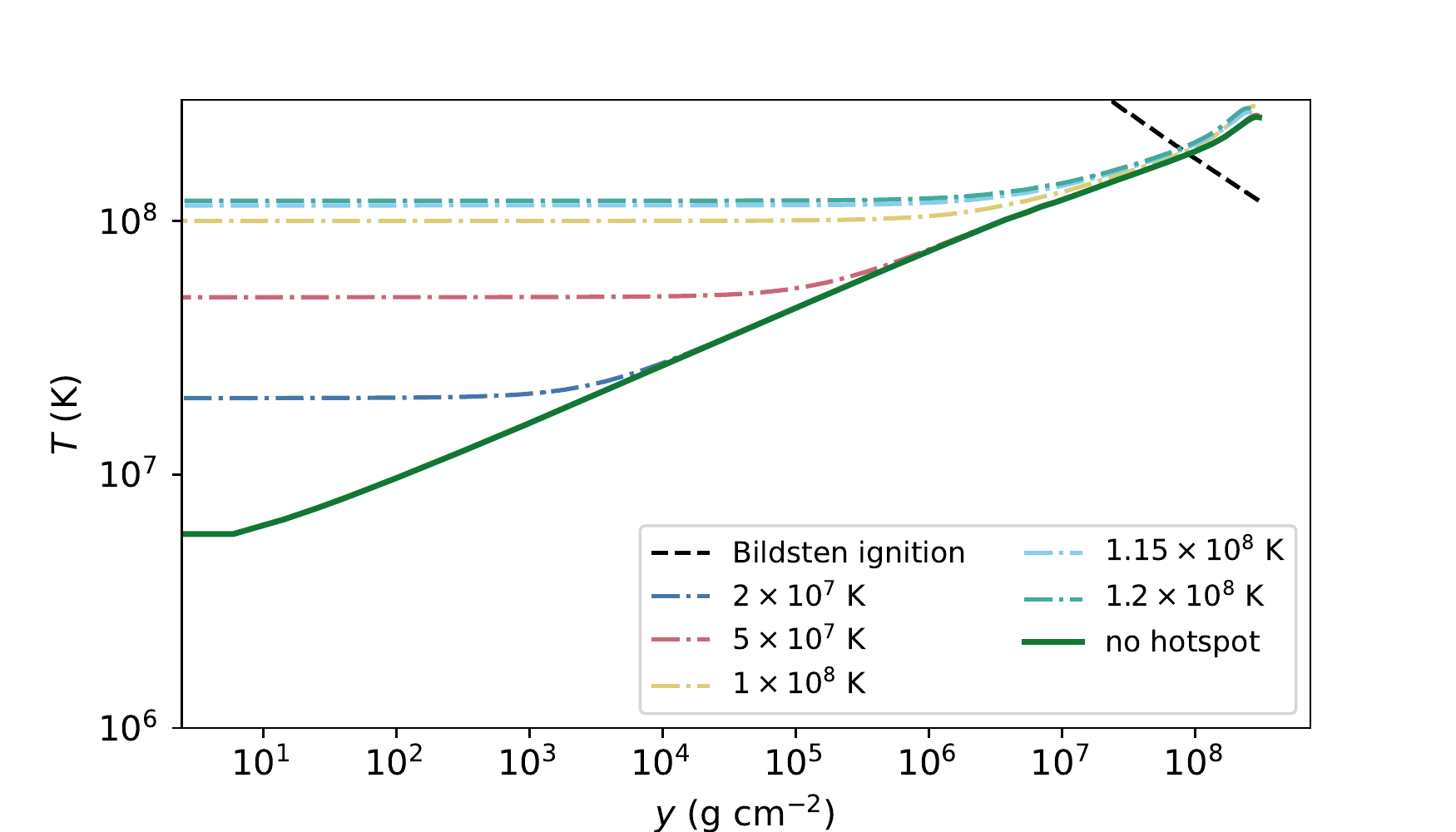}

    \caption{
    \textit{Left panels:} Individual radial columns under the hotspot for each temperature (dashed-dotted lines) and a radial column far from the hotspot location (solid line). The $T-y$ conditions for X-ray burst ignition (Eq.~\ref{eq:ignition}) are plotted (dashed black line). Since the temperature profile is not different from the KEPLER run without hotspot near the ignition depth, it is clear that ignition is not reached in any of the models, as expected since these models are 1 minute before ignition.
    \textit{Right panels:} The temperature distribution at ignition for different hotspot temperatures for the \textsc{Kepler} calculations. 
    \textit{Top panels:} Model~A;
    \textit{Middle panels:} Model~B;
    \textit{Bottom panels:} Model~C.
    }
    \label{fig:ignition}
\end{figure*}

The 2D temperature distributions and temperature profiles for the \XTE\, models 1 minute before convection are shown in Figure~\ref{fig:ignition}.  We found that there was no difference in temperature at the burst ignition depth for $T\lesssim1.0\times10^8\,\mathrm{K}$.  According to the \citet{Bildsten1998} ignition formula (Eq.~\ref{eq:ignition}), in these cases ignition would be reached at all locations on the star at the same time, however, we note that this is a very simple formula and since we were able to obtain a stationary solution, runaway could not have commenced in these models. Qualitatively, in Figure~\ref{fig:ignition}, it is clear that the hotter the hotspot, the deeper the heating penetrates, and closer to a burst the hotspot could indeed provide enough heating to trigger burst ignition shallower than the rest of the star.

\subsection{The case of IGR J17480--2446}
The observed and inferred system parameters for \IGR{} are shown in Table~\ref{tab:1748obs}. Based on these observed parameters, we set up our model such that $R_{\mathrm{HS}}=0.5\,\mathrm{km}$, $\dot{m}=0.038\,$\medd\footnote{We adopt this accretion rate as it falls in the range of accretion rates consistent with the persistent luminosity of \IGR, as well as being consistent with the lowest accretion rate observed from \XTE.}, we explored a range of hotspot temperatures, and call the models with this set of parameters Model~B.

\begin{table}
	\centering
	\caption{
	Observed and inferred parameters for \IGR}
	\label{tab:1748obs}
	\begin{tabular}{llll} 
		\hline
		Parameter & Value & Units & Ref. \\
		\hline
		$P$& 11 & Hz & [1]\\
		$L_X^*$ & $(1.9-9.2)\times10^{37}$ & erg\,s$^{-1}$ & [1] \\
		$\dot{m}$& 0.01--0.14 &$\dot{M}_{\mathrm{Edd}}$ & Eq.~\ref{eq:Mdot} \\ 
		$R_{\mathrm{pc}}$ & 0.57 & km & Eq.~\ref{eq:r-polar-cap} \\
		$k_\mathrm{B}T$ (burst) & 1.75--2.5 & keV & [2] \\
		$k_\mathrm{B}T$ (non-burst) & 0.7--0.9 & keV & [1] \\
		$B$ & $\sim2\times10^8 - 2.4\times10^{10}$ & G & [1, 3] \\
		\hline
	\end{tabular}
	\small{\\$L_X$ is the persistent X-ray luminosity. $^*$This luminosity range encompasses values measured through the outbursts. {Ref.:} [1] \citet{Papitto2012}, [2] \citet{Chakraborty2011}, [3] \citet{Ootes2019}
	}
\end{table}

The time before convection at which no stationary solution was found for \textsc{Kepler} dumps sequentially closer to runaway is shown in Table~\ref{tab:ignitiontime} (Model~B) for different hotspot temperatures. We found that for the 2D code runaway only occured earlier for hotspot temperatures $T_\mathrm{HS}\gtrsim1\times10^8\,\mathrm{K}$, and we found that runaway occured at least 7 seconds earlier; considerably earlier than the comparable case for \XTE. Similarly, for the \textsc{Kepler} calculations with the same parameters, we found that a hotspot of just $T_\mathrm{HS}\gtrsim2\times10^7$\,K was sufficient to induce runaway $47.5\,\mathrm{s}$ earlier than the case with no hotspot. Again, this is considerably earlier than the comparable case for \XTE. 

The 2D temperature distributions and temperature profiles for \IGR{} are shown in Figure~\ref{fig:ignition}. Similar to the models for \XTE{} we found that there was no difference in temperature at the burst ignition depth  for any of the models for the \citet{Bildsten1998} ignition criterion, however this ignition criterion does not agree with the location of runaway for the \textsc{Kepler} models. Qualitatively, as for \XTE, in Figure~\ref{fig:ignition}, it is clear that the hotter the hotspot, the deeper the heating penetrates, and closer to a burst the hotspot could indeed provide enough heating to trigger burst ignition shallower than the rest of the star.

On comparison of the temperature profiles for \XTE{} and \IGR{} we find qualitatively very similar results for both accretion rates explored, excepting that for the lower accretion rate, the accreted layer is thicker and cooler, and ignition occurs deeper at cooler temperatures.  In both cases we required a hotspot temperature $\gtrsim1.0\times10^8\,\mathrm{K}$ to see any difference in the ignition depth between the column under the hotspot and the column with no heated area on the surface.

\subsection{More general models: Hotspot at magnetic pole vs hot-stripe at equator}
Here we consider the effect that the fast rotation of the pulsar might have on the ignition location at the equator compared to the magnetic poles. 

The time before convection at which no stationary solution was found for \textsc{Kepler} dumps sequentially closer to runaway is shown in Table~\ref{tab:ignitiontime} (Model~C) for different hotspot temperatures with reduced gravity and Cartesian geometry. Using the 2D code we found that runaway occured earlier than the case with no hotspot for the models with much lower hotspot temperatures than at the pole, with runaway occuring $1$--$10\,\mathrm{s}$ earlier than the no hotspot case for $T_{\mathrm{HS}}=2\times10^7\,\mathrm{K}$.  This is considerably lower than the hotspot temperature required for earlier runaway at the pole, of $T_{\mathrm{HS}}=1.15\times10^8\,\mathrm{K}$, however, if there was a hot-stripe along the equator due to accretion it is not clear how thick this stripe might be. Here we have set the stripe to just $1\,\mathrm{km}$ thick.  For the \textsc{Kepler} calculations with the same parameters, we found that runaway occurred earlier for all hotspot temperatures simulated, with runaway occurring 4.9\,s earlier for $T_{\mathrm{HS}}=2\times10^7$\,K.

\begin{figure}
\includegraphics[width=\columnwidth]{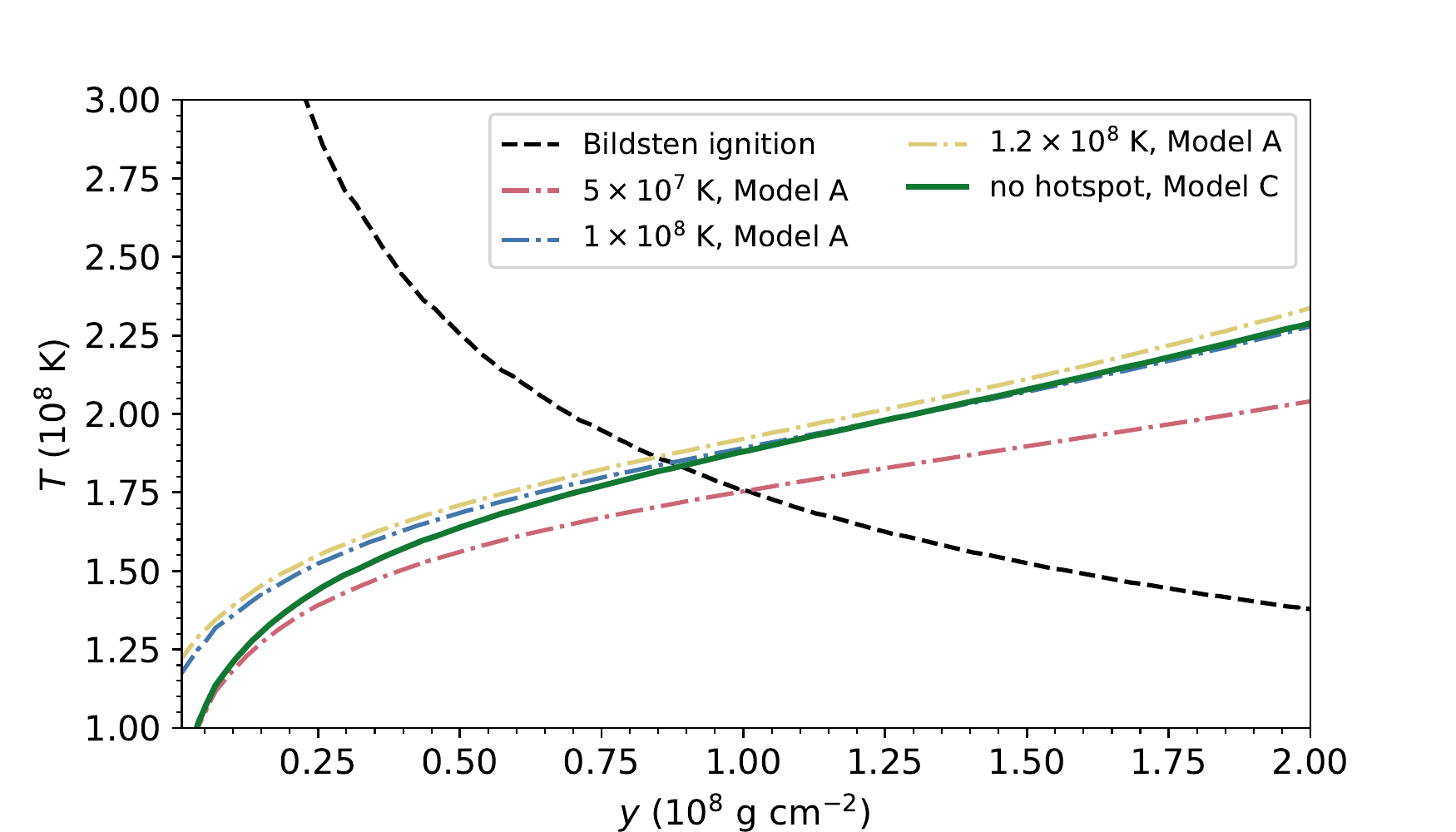}
    \caption{ Comparison of temperature profiles for a hotspot at the magnetic pole (dashed-dotted, Model~A) and the no hotspot case at the equator (solid line, Model~C) for the \textsc{Kepler} calculations. In our models the gravity at the equator is reduced by $25\,\%$ to account for the effect of fast rotation. The dashed black line shows the location of burst ignition (Eq.~\ref{eq:ignition}). Ignition is reached earlier (shallower) for Model~C at the equator except when the hotspot at the pole (Model~A) is $T_{\mathrm{HS}}\gtrsim1\times10^8\,\mathrm{K}$.}
    \label{fig:equatorcomparison}
\end{figure}

Next we consider the two sets of models, Model~A and Model~C (which are the same except Model~C has $25\,\%$ reduced gravity), and use cylindrical coordinates to simulate the pole for Model~A and Cartesian coordinates to simulate the equator for Model~C. We consider two hotspot temperatures at the pole, of $T_{\mathrm{HS}}=5\times10^7\,\mathrm{K}$ and $1\times10^8\,\mathrm{K}$, and set the radius of the hotspot to be 1\,km. The temperature profiles for these models are plotted in Figure~\ref{fig:equatorcomparison}. 

In Figure~\ref{fig:equatorcomparison} it is clear that if the star was uniformly at the same temperature, for rapidly rotating neutron stars (where we have assumed the surface gravity at the equator is reduced by 25\%) ignition would preferentially occur at the equator, as the temperature of the entire accreted column is slightly hotter (yellow line) than at the pole (red line). This result is in agreement with previous studies on ignition location by \citet{Spitkovsky2002,Cooper2007}. The effect of the hot-stripe at the equator is quantitatively the same as the effect of a hotspot at the pole, however due to the higher temperature at the equator, runaway occurs for a much colder hotspot. 

Interestingly, we found that for a $1\times10^8\,\mathrm{K}$ hotspot, ignition is likely reached underneath the hotspot at the pole before the equator (with no hotter area). In Figure~\ref{fig:equatorcomparison}, the temperature profiles are plotted 2 minutes before a burst (when convection begins in \textsc{Kepler}), and closer to the burst all columns would be hotter, and ignition would be reached under the $1\times10^8$\,K hotspot prior to the no hotspot column at the equator. \emph{This suggests that for a sufficiently hot hotspot, ignition could occur away from the equator at the magnetic pole, despite the lower surface gravity at the equator.}

\subsection{Influence of hotspot size and geometries}
\label{sec:hsize}
We explored the effect that different hotspot sizes and the geometries of our models had on the diffusion and heating induced in deeper layers by the hotspot. For a very small hotspot, it is expected that the temperature of deeper layers would not be affected.  As the hotspot increases in size, the deeper layers will become increasingly more affected. If we increase the size of the hotspot above a certain radius, the temperature in the column below the center asymptotes to a constant value (set by the temperature of the hotspot).  Due to the 2D nature of our code, we were able to investigate the relevant hotspot sizes that induce this behaviour. We consider models of different hotspot sizes (from $1\,\mathrm{cm}$--$2\,\mathrm{km}$) for cylindrical and Cartesian geometries and with $T_\mathrm{HS}=8\times10^7\,\mathrm{K}$, and we set the length of the domain in $x$ to be 4 times larger than the hotspot size (we set the number of zones in $x$-direction to $200$ to ensure that we could measure the extent reached by dissipation).  
We note that for the purposes of analysing the effect that the model geometry has on the results we did not reduce the gravity at the equator by $25\,\%$ for these models.

\begin{figure*}
	\includegraphics[width=\columnwidth]{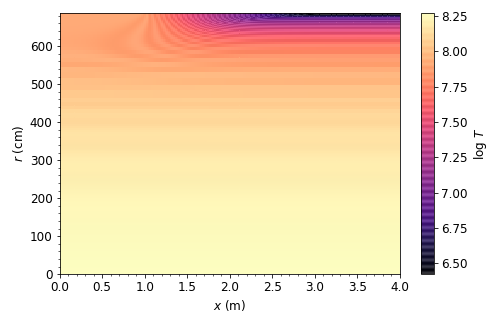}
	\includegraphics[width=\columnwidth]{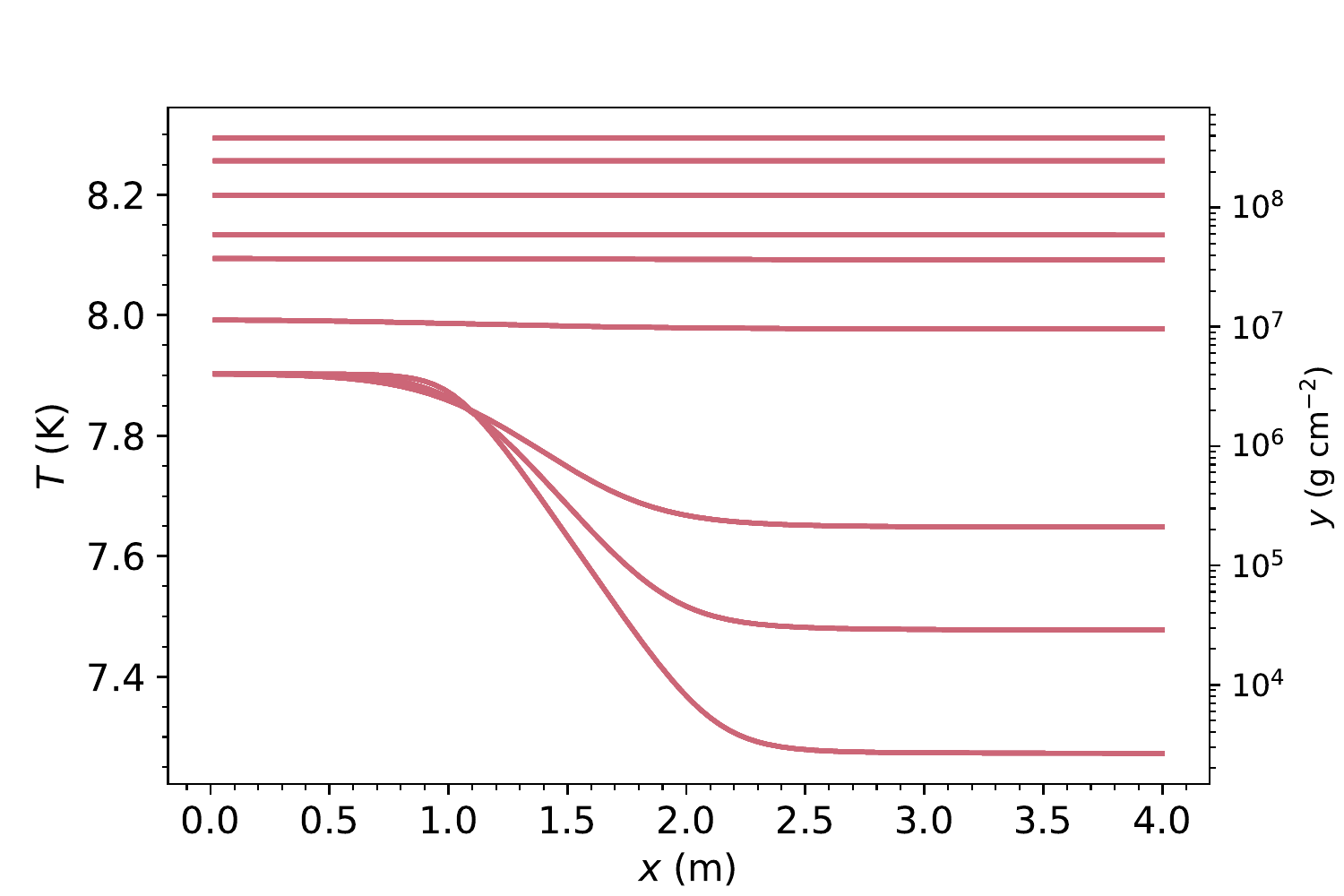}
	
	\includegraphics[width=\columnwidth]{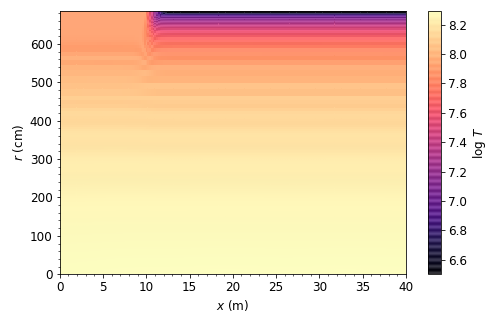}
	\includegraphics[width=\columnwidth]{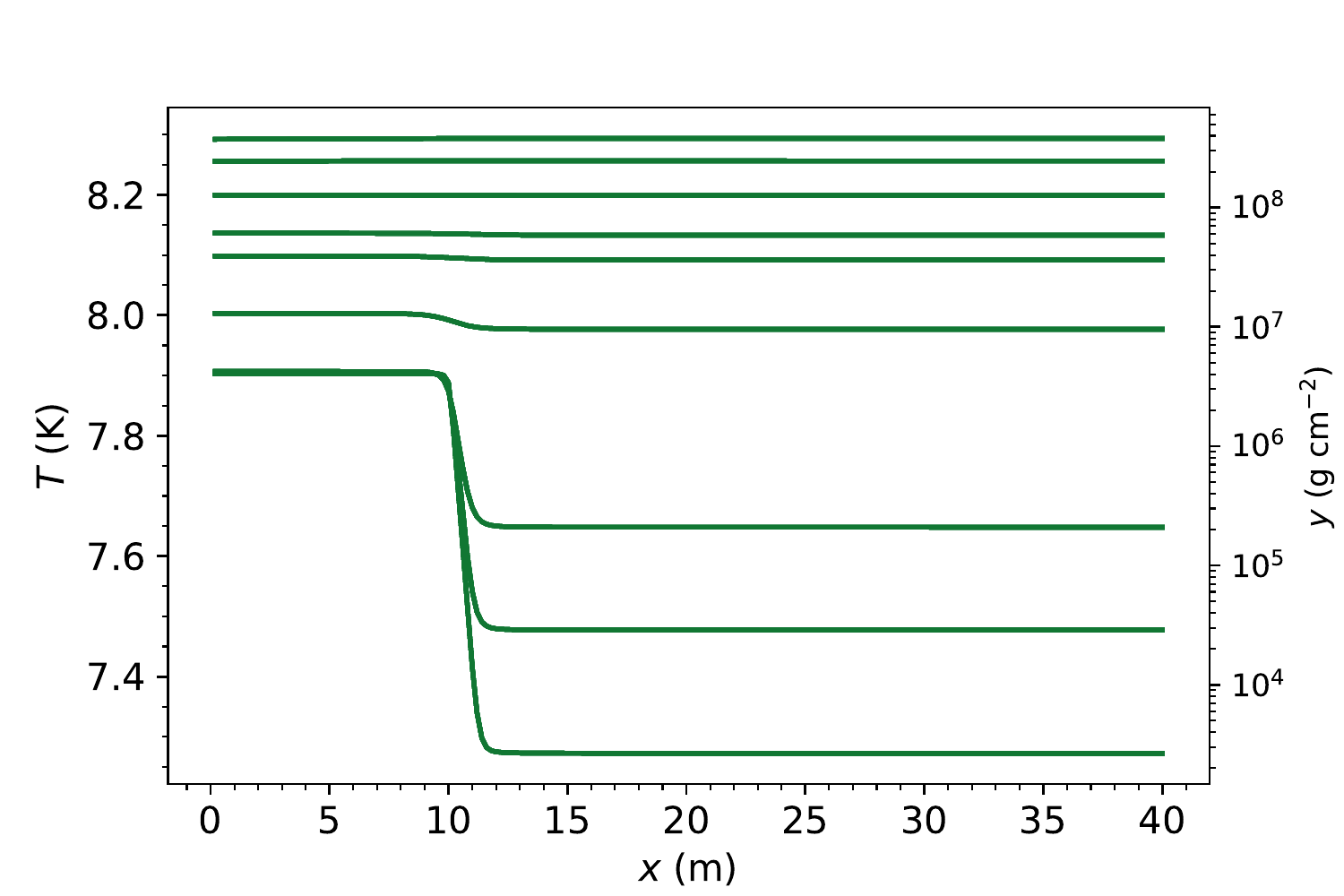}
	
	\includegraphics[width=\columnwidth]{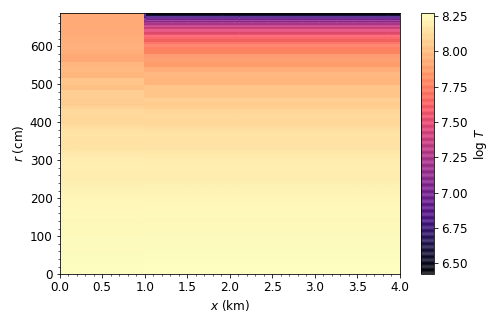}
	\includegraphics[width=\columnwidth]{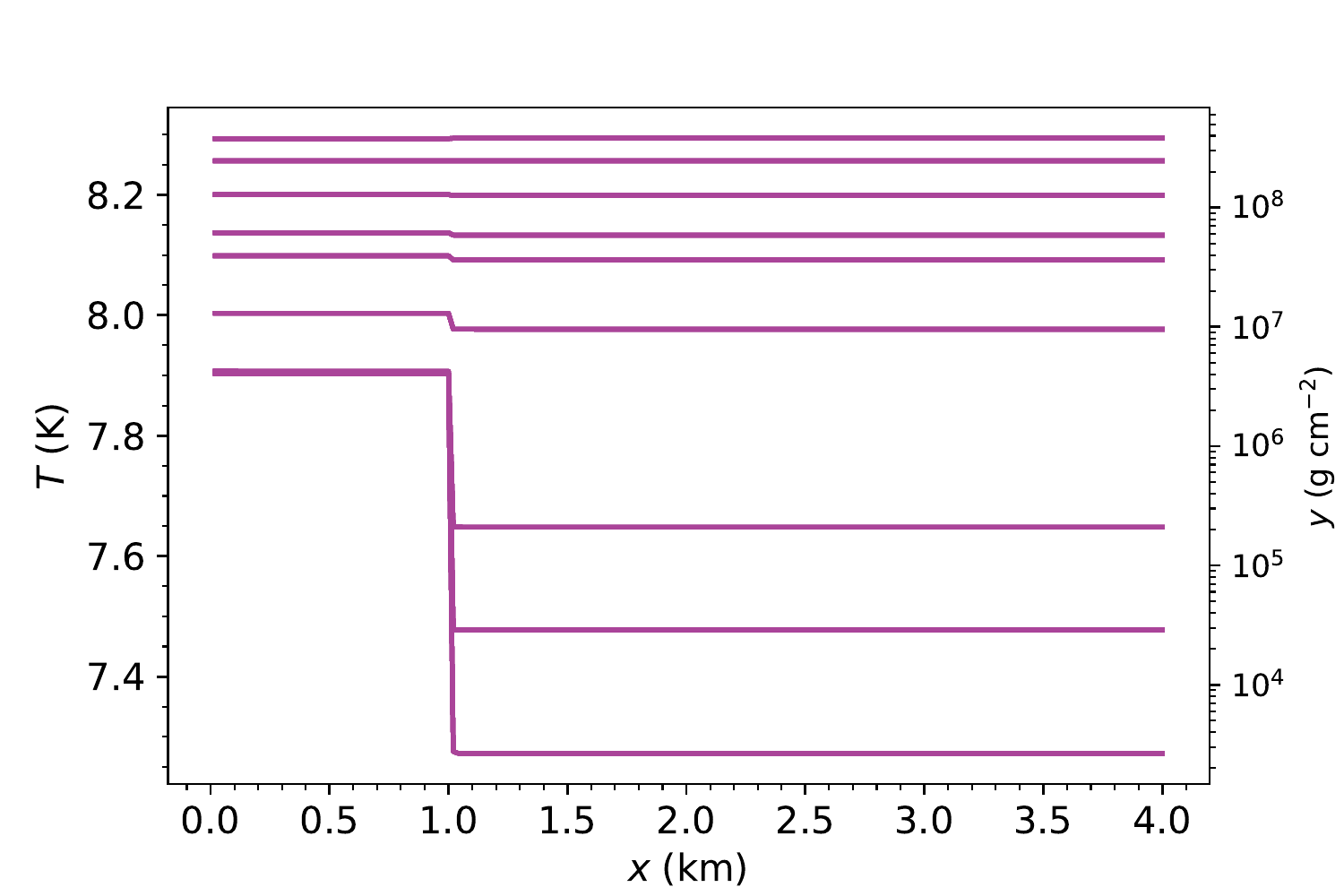}

    \caption{\textit{Top Left panel:} 2D temperature distribution for a hotspot radius of 1\,m with cylindrical geometry.  \textit{Top Right panel:} Temperature profiles for different radial depths for a hotspot of 1\,m showing how the heat from the hotspot disperses into the accreted layers. \textit{Middle Left panel:} 2D temperature distribution for a hotspot radius of 10\,m with cylindrical geometry.  \textit{Middle Right panel:} Temperature profiles for different radial depths for a hotspot of 10\,m showing how the heat from the hotspot disperses into the accreted layers. \textit{Bottom Left panel:} 2D temperature distribution for a hotspot radius of 1\,km with cylindrical geometry.  \textit{Bottom Right panel:} Temperature profiles for different radial depths for a hotspot of 1\,km showing how the heat from the hotspot disperses into the accreted layers.
    The model has parameters $\dot{m}=0.1$\,\medd, $Q_{\mathrm{b}}=0.1$\MeVnuc, and $T_{\mathrm{HS}}=8\times10^7$\,K, for a snapshot 1\,min before a burst begins. All temperatures are reported in log10.  }
    \label{fig:HS_2D_size}
\end{figure*}

The 2D temperature distribution and temperature profiles by radial depth for hotspots with radius 1\,m, 10\,m and $1\,\mathrm{km}$ and hotspot temperature $8\times10^{7}\,\mathrm{K}$ are plotted in Figure~\ref{fig:HS_2D_size}.  The 2D temperature distributions demonstrate how the heat diffuses outwards when the hotspots have smaller radii, which is not visible for the bigger hotspots due to the temperature saturation. For the $1\,\mathrm{m}$ case (top panels) it is evident that the temperature under the hotspot is not saturated, and horizontal diffusion is visible.  Due to the scale at which horizontal diffusion occurs, there is no such diffusion visible in the $1\,\mathrm{km}$ case (bottom panels). 

\begin{figure*}
	\includegraphics[width=\columnwidth]{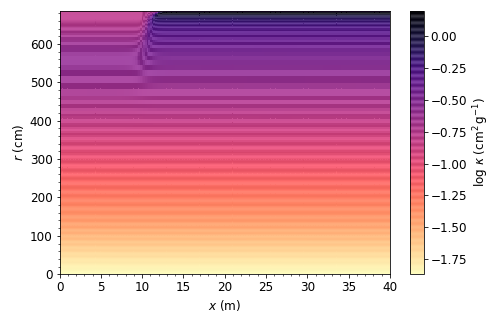}
	\includegraphics[width=\columnwidth]{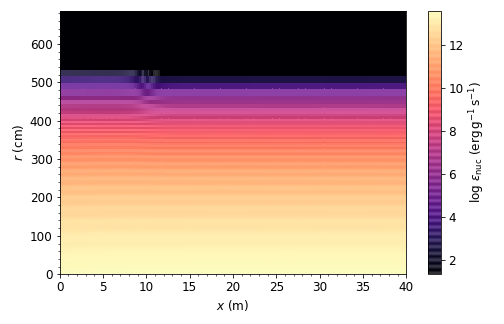}
     \caption{\textit{Left panel:} 2D opacity distribution for a hotspot radius of 10\,m with cylindrical geometry. \textit{Right panel:} 2D distribution of nuclear energy generation for a hotspot radius of 10\,m with cylindrical geometry.
    The model has parameters $\dot{m}=0.1$\,\medd, $Q_{\mathrm{b}}=0.1$\MeVnuc, and $T_{\mathrm{HS}}=8\times10^7$\,K, for a snapshot 1\,min before a burst begins. }
    \label{fig:HS_2D_size_kapeps}
\end{figure*}

In Figure~\ref{fig:HS_2D_size_kapeps} the 2D distributions of opacity and nuclear burning are plotted for a hotspot size of 10\,m. The opacity at the bottom of the accreted layer is very low ($\sim10^{-2}\,\mathrm{cm}^2\,\mathrm{g}^{-1}$), and the nuclear burning is very low near the surface, and comes mostly from the deeper accreted layers.

\begin{figure}
\includegraphics[width=\columnwidth]{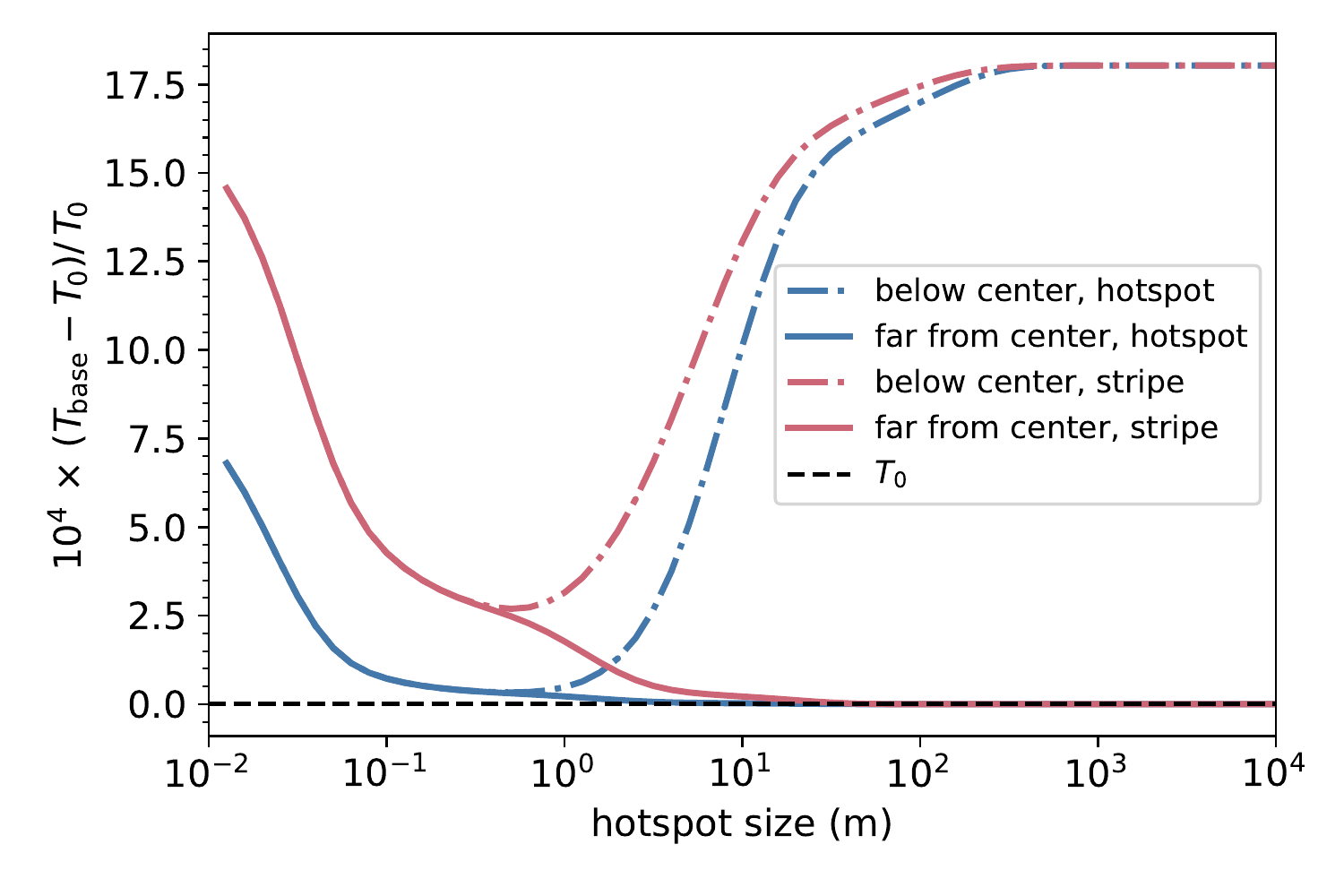}
\caption{The temperature at the base of the accreted layer for a column under the hotspot (dashed-dotted line) and one far from the hotspot (solid line), for cylindrical (red) and Cartesian (blue) geometries and a range of hotspot sizes (in 0.1\,dex steps). $T_0$ is the temperature at the base for a simulation with no hotspot. The model has parameters of Model~A ($\dot{m}=0.1$\,\medd, $Q_{\mathrm{b}}=0.1$\MeVnuc), and $T_{\mathrm{HS}}=8\times10^7$\,K, for a snapshot 2\,min before a burst begins in the \textsc{Kepler} simulation. Note that in this case we have not reduced the gravity in the Cartesian coordinate system simulations in order to analyse the effect of geometry alone. For hotspot sizes $\lesssim1\,\mathrm{m}$ the domain is so small that the heat from the hotspot spreads evenly by the time it reaches the bottom. This is an artifact of the code, but it would not be possible to resolve for the smallest sizes. We find the temperature under the hotspot asymptotes to a constant value for hotspots $\gtrsim100\,\mathrm{m}$.}
\label{fig:HSbottomtemp}
\end{figure}

In Figure~\ref{fig:HSbottomtemp} we have plotted the temperature at the bottom of the accreted layer ($T_{\mathrm{base}}$) for all hotspot sizes and the two geometries for the stationary solution 2\,min before a burst, in hotspot size steps of 0.1\,dex. 
Figure~\ref{fig:HSbottomtemp} demonstrates  that for hotspot sizes less than 1\,m, the heating does not penetrate to the bottom of the accreted layer, as there is no difference in temperature under the hotspot (dashed-dotted lines) and not under the hotspot (solid lines). Between 1\,m and 100\,m the temperature near the bottom of the accreted layer under the hotspot gradually becomes saturated. For hotspots bigger than $\sim$100\,m, i.e., \emph{hotspots much larger than the layer thickness}, the centre temperature of the hotspot is saturated, and the boundary between the hotspot and the cooler columns drops on a scale shorter than the hotspot size.  This, however, is not as trivial a result as it may naively seem.  This is because the opacity in the deeper accretion layers and in the substrate is orders of magnitude smaller than at the surface (Figure~\ref{fig:initial}), and hence heat could spread rather efficiently.  In fact, in some preliminary time-dependent simulations not further discussed here, we have seen a broad runaway up to several times the layer height for hotspots with widths of the order of the layer height.  Realistic hotspots, however, have sizes of $\sim1\,\mathrm{km}$ (see below), i.e., a factor 200 larger than the layer height,  so that the centre of the hotspot is no longer affected by its boundary.

For a hot-stripe, simulated in Cartesian plane-parallel geometry, the accreted layers are consistently hotter as the ratio of the perimeter to the interior is smaller in the case of a stripe, and heat can less easily escape. We found, however, that the relevant size at which the hot-stripe could affect deeper layers is very similar to the cylindrical coordinate case.  The centres of wide stripes hence behave the same way as the centre of large hotspots.

It is very unlikely that a hotspot caused by magnetic channelling in an AXP could be less than 10\,m in size, given the size of the polar cap (Eq.~\ref{eq:r-polar-cap}).  Therefore we are always operating in the regime where the accreted column under the hotspot is saturated, and heat diffusion horizontally from the hotspot is not important.  Thus, both in the case of a wide hotspot and a wide equatorial stripe the centre can be well studied using a 1D code such as \textsc{Kepler}.

\section{Discussion}\label{sec:discussion}
Our findings reveal that a hotspot on the surface of an accreting pulsar could influence the ignition location of X-ray bursts if the hotspot has a temperature $\gtrsim2\times10^7$\,K. We reproduce previous studies that find that due to the lower surface gravity expected at the equator, ignition will almost always occur at this location. We find that hotspots with a temperature less than $1\times10^8$\,K are unlikely to affect ignition location, but hotspots hotter than this could induce burst ignition at the magnetic pole, even when accounting for the expected lower surface gravity at the equator. Furthermore, when magnetic channelling is in place we do not expect a hot stripe to develop at the equator as in the case of pure disc accretion, so that ignition at the magnetic pole is even more likely.

Our results suggest that a hotspot with temperature $\gtrsim1\times10^8$\,K is required for X-ray burst ignition location to move from the equator to the magnetic pole. Whether hotspots on the surface can actually get this hot is uncertain, especially since spectral modelling of the X-ray emission from AXPs usually infers temperatures in the range 1--5$\times10^7$\,K or less. However, there is some uncertainty in the hotspot temperatures obtained via this method, as it can be difficult to separate the thermal hotspot emission from other thermal emitting regions in the system, which, however, are characterised by a colder temperature than the former.

An alternative (whilst simple) method of obtaining the hotspot temperature is by estimating the blackbody temperature of an emitting area with radius $R_{\mathrm{HS}}$, radiating with the accretion luminosity $L_{\mathrm{acc}}$ \citep{Frank2002}
\begin{equation}
    T_{\mathrm{bb}} = \left(\frac{L_{\mathrm{acc}}}{4\pi R_{\mathrm{HS}}^2 \sigma}\right)^{1/4}
\end{equation}
where $\sigma$ is the Stefan-Boltzmann constant. 

Assuming an accretion luminosity of $1\times10^{38}$\,erg and $R_{\mathrm{HS}}=0.5$\,km this gives $T_{bb}\approx8.7\times10^7$\,K. Alternatively, we could consider the thermal energy released if the gravitational potential energy were entirely converted into thermal energy of the motion of the electrons \citep{Frank2002}
\begin{equation}
    T_{\mathrm{th}} \approx \left( \frac{GMm_{\mathrm{p}}}{3\,k_\mathrm{B}\,R_{\star}} \right)
\end{equation}
where $m_{\mathrm{p}}$ is the mass of a proton, $M$ is the mass of the neutron star, and $R_{\star}$ is the radius of the neutron star. For a $1.4\,\msun$ neutron star and an emitting area with radius $0.5\,\mathrm{km}$, we find $T_{\mathrm{th}}\approx2.4\times10^8\,\mathrm{K}$. 

Thus the likelihood that the hotspot in \XTE, \IGR, or AXPs in general could be as hot as $1\times10^8\,\mathrm{K}$ is uncertain. It is expected that the true radiation temperature should lie between the blackbody temperature and the thermal temperature \citep{Frank2002}. In spectral modelling of \IGR, \citet{Papitto2012} found blackbody temperatures of the seed photons of 0.8--0.9\,keV, and temperatures as high as $3.5\times10^8\,\mathrm{K}$ for the electrons that Compton up-scatter the blackbody seed photons.
A hotspot with temperature $1\times10^8\,\mathrm{K}$ would be clearly evident in the X-ray spectrum as an 8.6\,keV blackbody component, which has never been conclusively observed in any accreting systems that we are aware of. We thus find it very unlikely that the hotspot in AXPs could be as hot as $1\times10^8\,\mathrm{K}$. 

If a hotspot can be as hot as $1\times10^8\,\mathrm{K}$, the scenario in which bursts ignite at the magnetic pole could be a plausible explanation for the burst oscillations observed in \IGR{} and \XTE{} that are coincident with the spin period of the pulsar in each of these systems. \citet{Cavecchi2011} suggested that the burst oscillations at the same frequency as the spin period of the pulsar in \IGR{} could be explained by a hotspot confined by hydromagnetic stresses to the magnetic pole.  They argue that due to tension induced by the compression of magnetic field lines, the strength of the magnetic field required for magnetic confinement is only $\sim4\times10^9\,\mathrm{G}$ \citep{Heng2009}.  This, combined with the ability to preferentially ignite X-ray bursts under the hotspot could well be a plausible explanation for these puzzling observations of \IGR.  The case of \XTE{}, however, is a little different, as the magnetic field in this system does not appear to be as strong as in \IGR. Observations cannot rule out a magnetic field as strong as $4\times10^9\,\mathrm{G}$ \citep{Bhattacharyya2005}, however, the magnetic field must be $\lesssim10^9\,\mathrm{G}$ in order for the magnetic propeller effect not to disrupt accretion \citep{Rappaport2004}. \citet{Watts2008} deduced that in the scenario in which magnetic confinement of the fuel is not possible, the effect of the higher temperature at the accretion impact point may induce burst ignition at this point.  In this case, in order to explain the continued phase locking of the burst oscillations with the accretion hotspot, the burning front of the burst would need to stall, perhaps due to the rate of heat transfer of the burning front.  In this work we deduce that it is plausible that the temperature asymmetries due to the accretion hotspot could cause bursts to ignite at the hotspot, however we need to include hydrodynamics and time-dependence in our simulations in order to assess if burst front stalling is possible. Regardless, the fact that the phase-locking of the burst oscillations with the persistent pulsations in \XTE{} happen only when the accretion rate is inferred to be higher \citep{Watts2008}, supports our finding that only when the hotspot is sufficiently hot can bursts ignite under the hotspot. 

We note that the persistent flux (and thus accretion rate) ranges listed in Table 2 and 3 for the case studies represent the absolute minimum and maximum constraints obtained from observations of these systems, encompassing the flux observed throughout the outbursts. In reality the best flux measurement for our modelling would be immediately prior to the onset of a burst. Nevertheless, we find that our models are not sensitive to the choice of accretion rate. This is because for higher accretion rates the whole column becomes hotter. Then, the time it takes to disperse the heating deeper into the surface depends solely on the diffusion rate, which is constant, and it is this diffusion time which determines whether ignition at the pole can overtake the equator.

In the results we report there are significant discrepancies between the 1D \textsc{Kepler} calculations of the influence of the hotspot on the ignition time, and the 2D heat transport code we developed.  These discrepancies can be attributed to a couple of things. 
Firstly, the influence of thermal inertia is clearly very important in calculations such as these, and the time-independent 2D heat transport code cannot account for this. Secondly, heating causes advection, which the 2D heat transport code does not model as it does not include hydrodynamics. \textsc{Kepler} solves the full time-dependent equations of conservation of mass, momentum, and energy, and thus can track the movement of material as it is being heated.  We deduce that it is most likely for this reason that there is such a discrepancy between the static 2D heat transport code and the \textsc{Kepler} model predictions for the runaway times in Table~\ref{tab:ignitiontime}. Thirdly, the ignition time inferred using the 2D heat transport code is heavily dependent on the numerical accuracy of the code, as we define ignition to be when the code no longer converges to a static solution. Whilst we mitigated this effect as much as possible by using static solutions for slightly cooler hotspots as the initial guess temperature, there is still some uncertainty introduced due to the numerical accuracy.  In Section~\ref{sec:hsize}, however, we demonstrated the utility of the 2D heat transport code over \textsc{Kepler}, as it enables the horizontal diffusion length scale to be determined.  We find that the temperature diffuses horizontally on a length scale $\sim10\,\mathrm{m}$, significantly smaller than the expected radius of the hotspot based on the polar-cap size.  Thus we deduce that the 1D \textsc{Kepler} models of the runaway time are reasonable approximations to the 2D case, since the hotspots are sufficiently large that any horizontal diffusion is negligible. 

The 2D model of heat transport inside the accreted layers on the surface of a neutron star that we have developed is by no means exhaustive, and our results are dependent on numerous assumptions.

Firstly, we assumed that the accreted material is almost pure helium, and only accounted for helium burning via the triple-$\alpha$ reaction for the energy released due to nuclear burning.  Since it is thought that the triple-$\alpha$ reaction should be the main trigger of thermonuclear runaway at the onset of a burst \cite[e.g.,][]{Bildsten1998}, this assumption is the simplest case for the nuclear burning, but excludes energy released due to burning via other pathways, and would neglect heating due to burning of hydrogen to helium in case of helium-rich accretors.  In the latter case, the influence of the hotspot may be much reduced.  This case may be studied in the future. 

Secondly, we only solve for the steady-state solution and do not solve the equations including time-dependence. Under this assumption, our models cannot reach thermonuclear runaway as there is no static solution during runaway, and thus we can only model the conditions in the accreted layers up to just before a burst occurs. Indeed, on comparison with the equivalent time-dependent \textsc{Kepler} calculations, there is a discrepancy in the time of runaway for the different model cases. 

Finally, in this model we only address heat transport mechanisms and do not include hydrodynamical effects.  This assumption is perhaps the most important limitation of our model to discuss, as the effect of the temperature gradient on the movement of material in the accreted layer is important to consider. For example, the heated material may rise in the hotspot, inducing mixing in the hotter layers and movement of the fuel.  This would be complicated by composition gradients that could stabilise against rise of material.  Our static model cannot account for this, and it may have an effect on the prevalence of a temperature gradient in deeper layers, or the depth at which ignition conditions are reached first \citep[see, e.g.,][]{Malone2011,Malone2014}. 

In favour of our approach, however, is that in a hotspot we may expect strong magnetic fields, strong enough to funnel the accretion flow, and these same magnetic fields may also prevent fluid flow unless there was significant ambipolar diffusion.  For the other case, accretion in a stripe at the equator, Coriolis forces may prevent fluid motions away from the equator due to angular momentum conservation.  A detailed study of these effects is beyond the scope of this paper, but see \citet{Spitkovsky2002,Cavecchi2015}. Further to this point, our modelling of the hotspot temperature required to take ignition away from the equator to the magnetic pole assumes that the spreading time from the pole to the equator of the accreted fuel is long enough to allow ignition at the equator, or that accretion occurs both at the equator and the pole. If this were not the case, a slightly cooler hotspot may be enough to trigger ignition at the pole away from the equator. In fact, in the case of accreting millisecond pulsars, the fraction of mass accreted at the equator could be very low compared to the pole, and there may well be a viable scenario in which accretion primarily occurs at the pole and ignition could occur with a hotspot temperature significantly less than $10^8$\,K.

In future, the model should be upgraded to include hydrodynamical affects in order to conclusively examine the effect that a hotspot has on the deeper layers and location at which ignition conditions are reached first.
A more realistic nuclear reaction network that accounts for energy released by reactions other than the triple-$\alpha$ reaction and time-dependence should also be included so that the full thermonuclear runaway can be modelled. 

\section{Conclusions}\label{sec:conclusion}
In this study we have presented 2D calculations of X-ray burst ignition location on the surface of accreting pulsars, taking into account the effect of temperature asymmetries caused by a hotspot on their surface.
We found that heating persisted down to the approximate column depth of X-ray burst ignition for a hotspot temperature $\sim(2)\times10^7$\,K and independent of the size of the hotspot (assuming the hotspot radius is $\gtrsim100$\,m). In models of accretion at the equator and in models where the accreted fuel spreads evenly over the neutron star before ignition, we found that due to the lower surface gravity at the equator for rapidly rotating neutron stars, burst ignition will always preferentially ignite here unless the hotspot at the magnetic pole is hotter than $\gtrsim(1)\times10^8$\,K. At this temperature ignition would preferentially occur under the hotspot at the magnetic pole, providing a scenario in which off-equator ignition could occur in an accreting pulsar. However, due to lack of observational evidence, we find it unlikely that the hotspot in \XTE, \IGR, or any AXP could be as hot as $1\times10^8$\,K. We conclude that the ignition of bursts at the hotspot combined with magnetohydrodynamic effects leading to hotter and colder patches associated with the magnetic pole (for example hydromagnetic modes) could explain the phase locking of burst oscillations with the accretion-powered pulsations in the two AXPs \XTE{} and \IGR. Whilst this scenario is promising, in order to comprehensively assess if it is possible to ignite at the hotspot, we would need to include a full treatment of hydrodynamics in the 2D heat transport code, time-dependence in our simulations, as well as a more robust nuclear reaction network. Our modelling demonstrates that the heat diffusion from the hotspot alone is unlikely to induce burst ignition at the magnetic pole, as it would require hotspot temperatures significantly greater than those observed in the X-ray spectra of accreting neutron stars.

\section*{Acknowledgements}

The authors thank Duncan Galloway, Bernhard M\"uller and Andrew Cumming for helpful discussions, and the anonymous referee for helpful and constructive comments. AJG acknowledges support by an Australian Government Research Training Program scholarship. This work benefited from support, in part, by the National Science Foundation under Grant No. PHY-1430152 (JINA Center for the Evolution of the Elements); by the Australian Research Council Centre of Excellence for Gravitational Wave Discovery (OzGrav), through project number CE170100004; by the Australian Research Council Centre of Excellence for All Sky Astrophysics in 3 Dimensions (ASTRO 3D), through project number CE170100013; and from discussions at the Lorentz center workshop \emph{"Bursting the Bubble: Connecting Thermonuclear Burst Research to a Wider Community"}\/ held in Leiden, June 2019. FRNC and ALW acknowledge support from ERC Starting Grant No. 639217 CSINEUTRONSTAR (PI Watts).

This research made use of Matplotlib, a community-developed core \textsc{python} package \citep{Matplotlib} and NASA’s Astrophysics Data System Bibliographic Services.

\section*{Data Availability}
The software written for the purpose of this work is publicly available on Github at \url{https://github.com/adellej/ignition-calculations}




\bibliographystyle{mnras}
\bibliography{bibfile} 

\bsp	
\label{lastpage}
\end{document}